\documentclass[prx,twocolumn,superscriptaddress]{revtex4-1}
\usepackage{amsmath}
\usepackage{amssymb}
\usepackage{amsthm}
\usepackage{amsfonts}
\usepackage{listings}
\usepackage{diagbox}
\usepackage{enumerate}
\usepackage{latexsym}
\usepackage{color}
\usepackage{xcolor}
\usepackage{bm}
\usepackage{hyperref}
\usepackage{graphicx}
\usepackage[FIGTOPCAP]{subfigure}
\usepackage{physics} 
\hypersetup{
    pdfnewwindow=true, colorlinks=true,
    linkcolor=blue, anchorcolor=blue,
    citecolor=blue, filecolor=blue,
    menucolor=blue, urlcolor=blue}

\newcommand{\nc}{\newcommand}
\nc{\webirvsp}{\href{https://github.com/zjwang11/irvsp}{\ttfamily irvsp}}
\nc{\webirtb}{\href{https://github.com/zjwang11/irvsp}{\ttfamily ir2tb}}
\nc{\webchecktopmat}{\href{https://www.cryst.ehu.es/cryst/checktopologicalmagmat}{\ttfamily Check Topological Mat}}
\nc{\weburl}{\href{http://link.aps.org/supplemental/10.1103/PhysRevB.105.224103}{\ttfamily http://link.aps.org/supplemental/10.1103/\\PhysRevB.105.224103}}
\nc{\bdel}{\vb* \delta}
\nc{\bsig}{\vb* \sigma}
\nc{\btau}{\vb* \tau}

\nc{\ba}{\vb* a}
\nc{\bb}{\vb* b}
\nc{\bc}{\vb* c}
\nc{\bk}{\vb* k}
\nc{\bp}{\vb* p}
\nc{\bs}{\vb* s}
\nc{\bx}{\vb* x}
\nc{\by}{\vb* y}
\nc{\bz}{\vb* z}

\nc{\bbI}{\mathbb I}
\nc{\calT}{\mathcal{T}}
\nc{\calI}{\mathcal{I}}

\nc{\hba}{\hat{ a}}
\nc{\hbb}{\hat{ b}}
\nc{\hbc}{\hat{ c}}
\nc{\hbx}{\hat{ x}}
\nc{\hby}{\hat{ y}}
\nc{\hbz}{\hat{ z}}

\nc{\mU}{{\mathcal U}}

\nc{\dg}{\dagger}
\nc{\ua}{\uparrow}
\nc{\da}{\downarrow}
\nc{\mto}{\mapsto}

\nc{\ie}{i.e., }
\nc{\eg}{e.g., }
\nc{\ea}{\textit{et al.}}
\nc{\qhc}{\:\text{H.c.}\:}

\nc{\eV }{\text{eV}}
\nc{\AAA}{\text{ \AA}}
\nc{\degree}[1]{${#1}^{\circ}$}
\nc{\red}[1]{\textcolor{red}{#1}}

\begin{document}
\tolerance 10000
\draft
\title{%
Twisted nodal wires and three-dimensional quantum spin Hall effect \\
in distorted square-net compounds
}

\author{Junze Deng}
\affiliation{Beijing National Laboratory for Condensed Matter Physics, and Institute of Physics, Chinese Academy of Sciences, Beijing 100190, China}
\affiliation{University of Chinese Academy of Sciences, Beijing 100049, China}

\author{Dexi Shao}
\email{sdx@iphy.ac.cn}
\affiliation{Beijing National Laboratory for Condensed Matter Physics, and Institute of Physics, Chinese Academy of Sciences, Beijing 100190, China}
\affiliation{University of Chinese Academy of Sciences, Beijing 100049, China}
\affiliation{Department of Physics, Hangzhou Normal University, Hangzhou 311121, China}

\author{Jiacheng Gao}
\affiliation{Beijing National Laboratory for Condensed Matter Physics, and Institute of Physics, Chinese Academy of Sciences, Beijing 100190, China}
\affiliation{University of Chinese Academy of Sciences, Beijing 100049, China}

\author{Changming Yue}
\affiliation{Department of Physics, University of Fribourg, 1700 Fribourg, Switzerland}

\author{Hongming Weng}
\affiliation{Beijing National Laboratory for Condensed Matter Physics, and Institute of Physics, Chinese Academy of Sciences, Beijing 100190, China}
\affiliation{University of Chinese Academy of Sciences, Beijing 100049, China}

\author{Zhong Fang}
\affiliation{Beijing National Laboratory for Condensed Matter Physics, and Institute of Physics, Chinese Academy of Sciences, Beijing 100190, China}
\affiliation{University of Chinese Academy of Sciences, Beijing 100049, China}

\author{Zhijun Wang}
\email{wzj@iphy.ac.cn}
\affiliation{Beijing National Laboratory for Condensed Matter Physics, and Institute of Physics, Chinese Academy of Sciences, Beijing 100190, China}
\affiliation{University of Chinese Academy of Sciences, Beijing 100049, China}

\date{\today}

\begin{abstract}
    Recently, square-net materials have attracted lots of attention for the Dirac semimetal phase with negligible spin-orbit coupling (SOC) gap, \eg ZrSiS/LaSbTe and CaMnSb$_2$.
    In this paper, we demonstrate that the Jahn-Teller effect enlarges the nontrivial SOC gap in the distorted structure, \eg LaAsS and SrZnSb$_2$.
    Its distorted $X$ square-net layer ($X=$ P, As, Sb, Bi) resembles a quantum spin Hall (QSH) insulator.
    Since these QSH layers are simply stacked in the $\hat{x}$ direction and weakly coupled, three-dimensional QSH effect can be expected in these distorted materials, such as insulating compounds CeAs$_{1+x}$Se$_{1-y}$ and EuCdSb$_2$.
    Our detailed calculations show that it hosts two twisted nodal wires without SOC [each consists of two noncontractible time-reversal symmetry- and inversion symmetry-protected nodal lines touching at a fourfold degenerate point],
    while with SOC it becomes a topological crystalline insulator with symmetry indicators $(000; 2)$ and mirror Chern numbers $(0, 0)$.
    The nontrivial band topology is characterized by a generalized spin Chern number $C_{s+}=2$ when there is a gap between two sets of $\hat{s}_{x}$ eigenvalues.
 The nontrivial topology of these materials can be well reproduced by our tight-binding model and  the calculated spin Hall conductivity is quantized to $\sigma^{x}_{yz} = (\frac{\hbar}{e})\frac{G_xe^2}{\pi h}$ with $G_x$ a reciprocal lattice vector.
\end{abstract}
\maketitle

\section{Introduction}
\begin{figure}[ht!]
    \centering
    \includegraphics[width=0.44\textwidth]{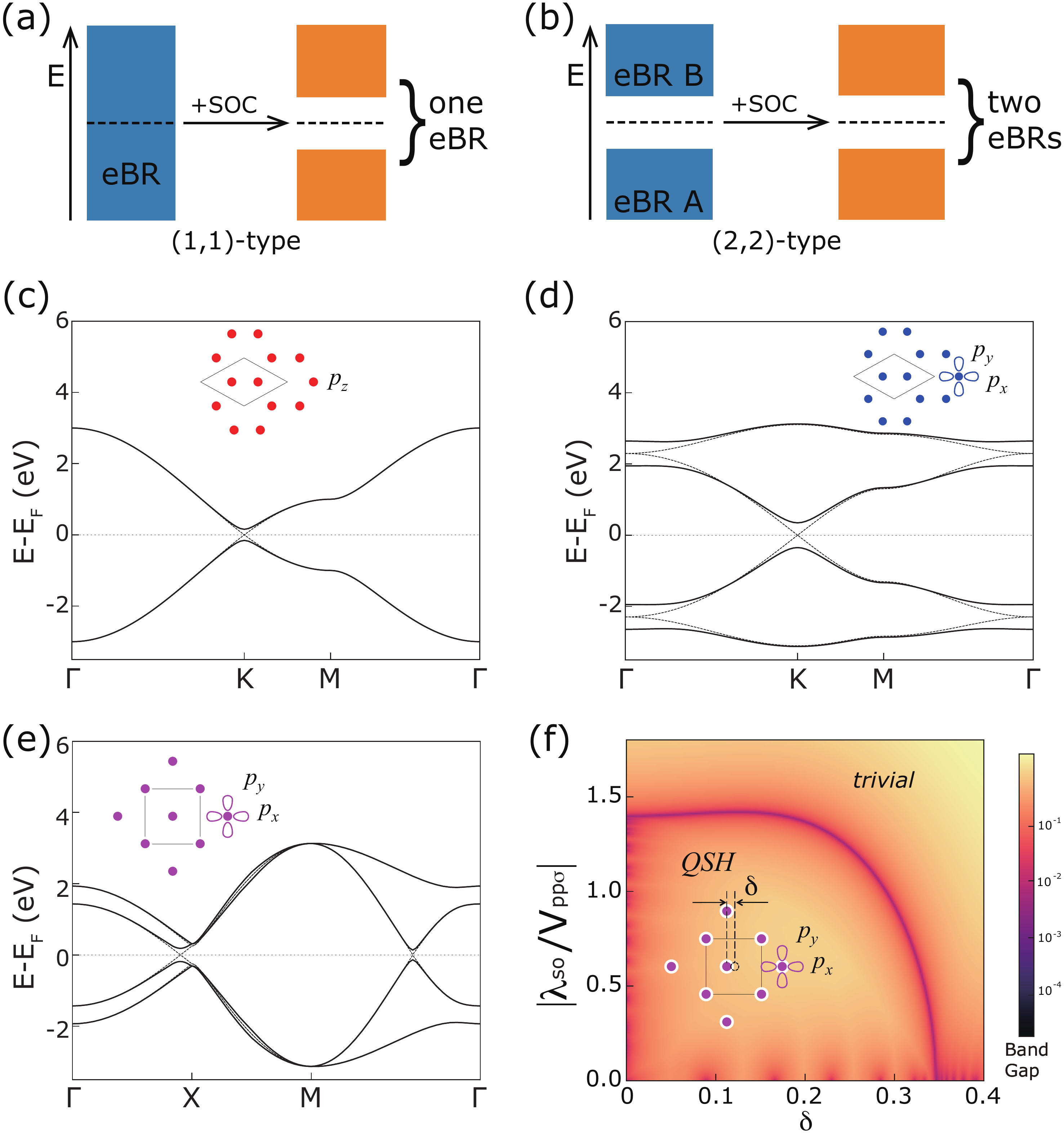}
    \caption{
        Topological phase transition induced by SOC in (a) $(1, 1)$ type and (b) $(2, 2)$ type.
        A brown block denotes an eBR, while a yellow block denotes a non-eBR.
        In (c)--(e) lattice models, dashed (solid) lines represent bands without (with) SOC.
        (c) $p_z$ orbitals on a honeycomb lattice of $(1, 1)$ type.
        (d) $p_{x,y}$ or $d_{xz,yz}$ orbitals on a honeycomb lattice of $(1, 1)$ type.
        (e) $p_{x,y}$ orbitals on a square lattice of $(1, 2)$ type.
        (f) The phase diagram of a distorted square net indicated by the band gap.
        }
    \label{fig:1}
\end{figure}

Over the past decade, topological materials \cite{TIs-Hasan2010, TIs-FuLiang2011, Armitage2018, BernevigTIandTSC, PhysRevResearch.3.L012028, PhysRevB.103.115145, PhysRevB.101.155143, GAO2021667} have intrigued many interests in both theory and experiment.
Among topological insulators (TIs), SOC plays an important role for the nontrivial energy gap.
In topological quantum chemistry \cite{Bradlyn2017, Bernevig-EBR-2018, Bernevig-band-graphtheory-2018, eletride2021}, when the phase transition is driven by SOC, we label the transition class by $(n, m)$,
where $n$ denotes the number of elementary band representations (eBRs) near the Fermi level ($E_F$) in the absence of SOC, while $m$ denotes the number of derived eBRs in the presence of SOC.
Therefore, without SOC, in the $(1, 1)$- or $(1, 2)$-type material [Fig. \ref{fig:1}(a)], its valence bands (VBs) and conduction bands (CBs) belong to an eBR (\ie a semimetal),
while in the $(2, 2)$-type material [Fig. \ref{fig:1}(b)], its VBs and CBs belong to two eBRs (\ie an insulator), \eg HgTe/CdTe quantum wells \cite{BHZScience, BHZScienceExp, PhysRevB.72.035321BHZpara}, and 3D TI Bi$_2$Se$_3$ \cite{Bi2X3-zhj2009, Hsieh2009, Bi2X3-Hasan2009, Bi2Te3-Shen2009}.
When including SOC and varying its strength ($\lambda_{\text{so}}$), band inversion occurs in the $(2, 2)$ type ($\lambda_{\text{so}}> \lambda_{c}$); namely, the topological phase transition is accidental.
However, in the $(1, 1)$ and $(1, 2)$ types, the gapped phases driven by infinitesimal SOC are topologically nontrivial, corresponding to a topological metal-insulator transition (TMIT) without \emph{band inversion}.

Here are several examples of TMIT.
A well-known one is graphene (or silicene), where the Dirac semimetal phase originates from the half filling of the eBR $A_1@2b$ (SG 183; $p_z$ orbital) without SOC.
Infinitesimal SOC gaps its Dirac point at $K$ [Fig. \ref{fig:1}(c)] and makes it a quantum spin Hall (QSH) insulator \cite{PhysRevLett.95.226801KaneMeleRef1, PhysRevLett.95.146802KaneMeleRef2, PhysRevLett.107.076802}.
Another is bismuthene on a SiC substrate \cite{PNAS.140970111SiCRef1, Hsu_2015SiCRef2, Reis287}, which was proposed in the original theoretical works \cite{PNAS.140970111SiCRef1, Hsu_2015SiCRef2}.
The $p_z$ states are far below $E_F$ due to its strong coupling with the substrate, while the low-energy bands near $E_F$ forms the eBR $E@2b$ (SG 183; $p_{x,y}$ orbitals) with half filling.
It becomes a QSH insulator with SOC as well [Fig. \ref{fig:1}(d)].
There are more examples hosting arbitrary SOC induced TMIT without involving band inversion, \eg flat-band kagome systems \cite{PhysRevB.78.125104FlatBandTPTRef1, PhysRevLett.124.183901FlatBandTPTRef2, PhysRevB.105.085128FlatBandTPTRef3}.
In this paper, we have investigated the family of square-net materials, which attract lots of attention since the discovery of the anisotropic Dirac fermions in Ca/SrMnBi$_2$ \cite{PhysRevB.103.125131, 2019Klemenz, apl112111, PhysRevB.87.245104}.
Recently, a series of experimental progresses on quantum transport have been reported \cite{Liu2017, Liu2021, PhysRevB.100.195123,pnas.1706657114}.
In an $X$ square-net compound, the key feature of its band structure (BS) is the half-filling eBR $E@2a$ (SG 129; $p_{x,y}$ orbitals).
The SOC effect leads an square-net layer into QSH phase [Fig. \ref{fig:1}(e)].
Unfortunately, the SOC-induced topological gap is usually rather small in these compounds \cite{PhysRevB.92.205310}.

In this paper, we find that in distorted square-net compounds of $MXZ$ (LaAsS family) and $ABX_2$ (SrZnSb$_2$ family) \cite{HULLIGER1977371, Strauss2003CrystalSO, PhysRevB.87.245104}, the Jahn-Teller effect enlarges the nontrivial gap, which reduces the density of states at $E_F$.
We propose these compounds with distorted $X$ square-net layers resemble three-dimensional (3D) QSH effect \cite{Wang2016Hourglass}.
In the absence of SOC, the system is a nodal-line semimetal with two twisted nodal wires.
Each nodal wire consists of two noncontractible time-reversal symmetry ($\calT$) and inversion symmetry ($\calI$) -protected nodal lines touching at a fourfold degenerate point protected by $\tilde{C}_{2y}$ and $\calT\tilde{C}_{2z}$.
Once including SOC, it becomes a topological crystalline insulator (TCI) \cite{PhysRevLett.106.106802TCI, TCIExp, annurev-conmatphys-031214-014501} with symmetry indicators (SIs) $(z_2z_2z_2;z_4)=(000;2)$ and $M_y$ mirror Chern numbers (MCNs) $(m_0,m_\pi)=(0,0)$.
With a spectrum gap between two sets of $\hat s_x$ eigenvalues, its nontrivial nature is characterized by a generalized spin Chern number (SCN) $C_{s\pm} = \pm 2$.
The 3D QSH effect shall be expected in samples of insulating candidates, such as CeAs$_{1+x}$Se$_{1-y}$ and EuCdSb$_2$.
The nontrivial topology can be well reproduced by our tight-binding (TB) model and  the calculated spin Hall conductivity (SHC) is quantized, \ie $\sigma^{x}_{yz}= (\frac{\hbar}{e})\frac{G_xe^2}{\pi h}$ (with $G_x$ a reciprocal lattice vector).

\begin{figure}[!t]
    \centering
    \includegraphics[width=0.45\textwidth]{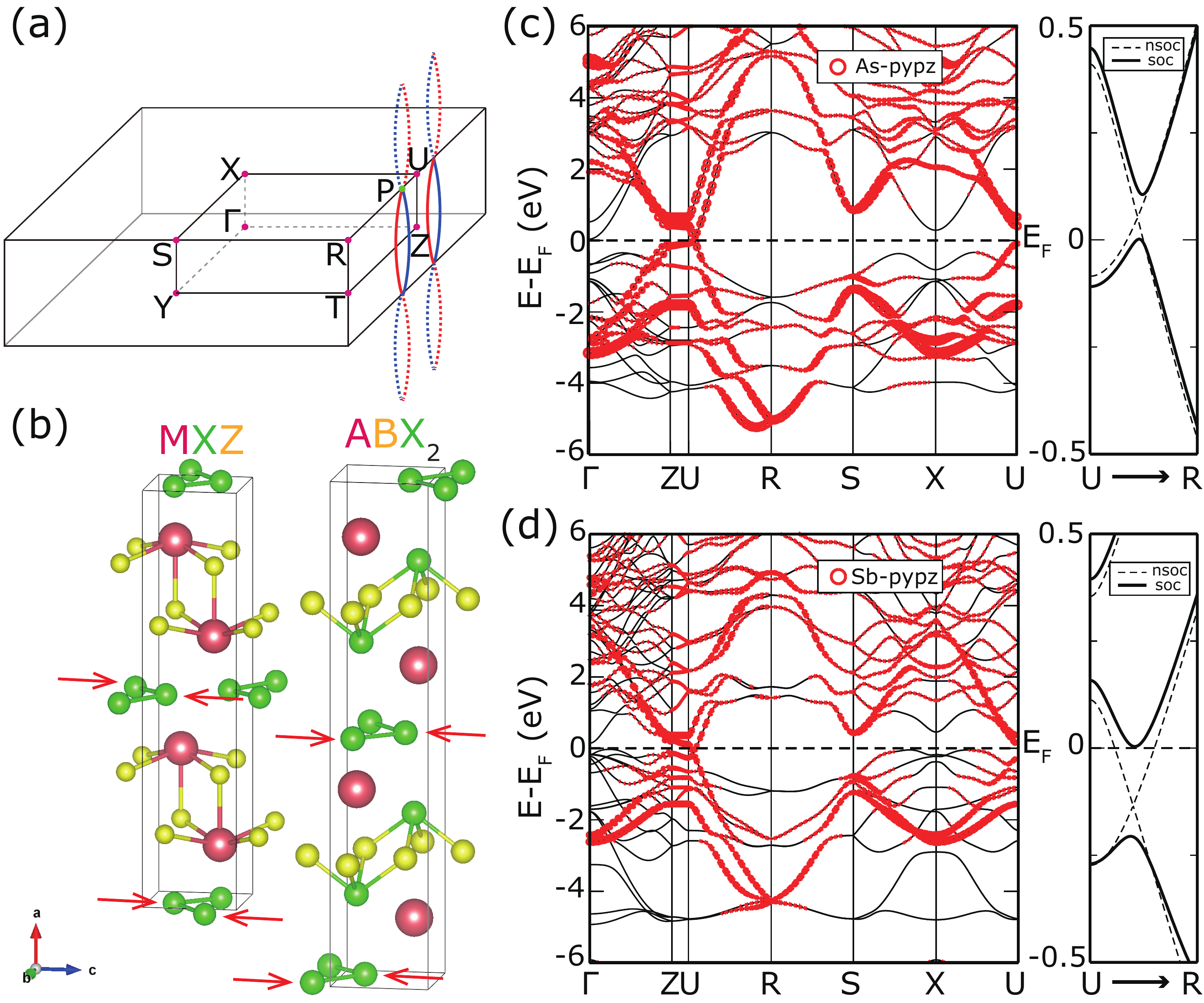}
    \caption{
        The Brillouin zone (BZ), crystal structures, and BSs of the $MXZ$ and $ABX_2$ compounds.
        (a) The bulk BZ with two twisted nodal wires colored in red and blue.
        Each wire consists of two noncontractible $\calT\calI$-protected nodal lines touching a fourfold degenerate point P (green point on U--R).
        (b) Crystal structures of the corresponding compounds. Each unit cell contains two distorted $X$ square-net layers. The red arrows illustrate the key distortion.
        (c) Orbital-resolved BSs of $Pnma$ PrAsS without SOC.
        (d) Orbital-resolved BSs of $Pnma$ EuCdSb$_{2}$ without SOC.
        The right panels of (c) and (d) indicate the band crossings along U--R disappear when SOC is taken into consideration.
        }
    \label{fig:2}
\end{figure}

\section{Calculations and Results}
\subsection{Crystal structures}
\begin{figure*}[ht!]
    \centering
    \includegraphics[width=0.95\textwidth]{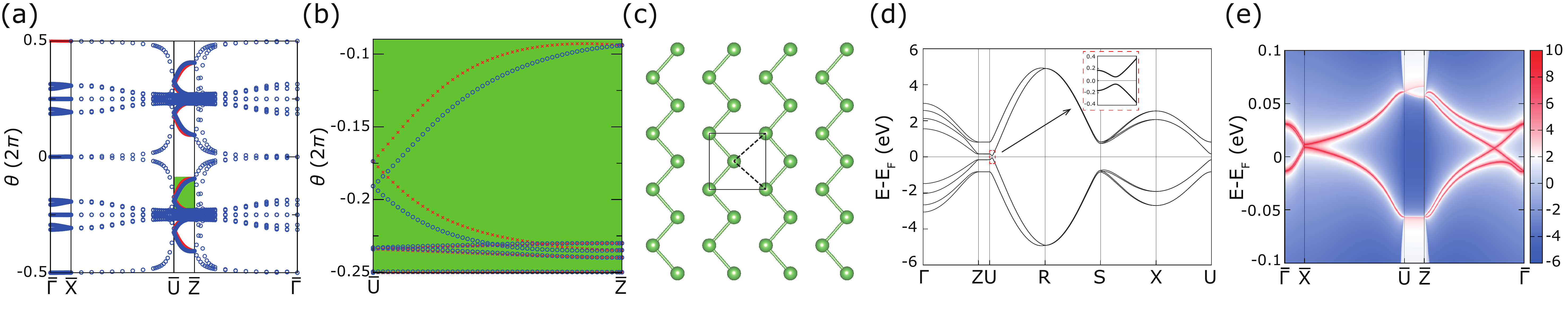}
    \caption{
        (a) The $k_y$-directed Wilson-loop spectrum for PrAsS along $\rm{\bar{\Gamma}\bar{X}\bar{U}\bar{Z}\bar{\Gamma}}$ in the (010)-surface BZ.
        (b) Close-up of the green region in (a); one can find the hourglass-shaped pattern of the Wilson-loop spectrum along $\rm{\bar{U}\bar{Z}}$, the crossings and circles denotes different $g_{x}\equiv \mathcal{I}\tilde{C}_{2x} \equiv \{\mathcal{I}C_{2x}|1/2,1/2,1/2\}$ eigenvalues.
        (c) The illustration of distorted square net.
        (d) BS with SOC of the minimum TB model, the inset is the close-up of BS around U.
        (e) (010)-surface spectrum of the minimum TB model. The hourglass-shaped surface states are presented on $\rm{\bar{\Gamma}\bar{X}}$.
        }
    \label{fig:3}
\end{figure*}

The $MXZ$ and $ABX_2$ compounds host an orthorhombic lattice (parameters $a,b,c$ along $\hbx,\hby,\hbz$, respectively), which is a distorted structure from SG 129 (doubling the unit cell in the $\hbx$ direction).
As illustrated in Fig. \ref{fig:2}(b), each unit cell contains two $X$ layers ($x=0$ and $0.5a$), which are slightly distorted square nets ($yz$ planes, parametrized by the displacement $\delta~c$ in the $\hbz$ direction).
Thus, the zigzag chains are formed along $\hby$ in the plane [Fig. \ref{fig:3}(c)], resulting in the $X^{1-}$ state.

\subsection{Twisted nodal wires in the absence of SOC}
From the orbital-resolved BSs of paramagnetic PrAsS and EuCdSb$_2$ in Figs. \ref{fig:2}(c) and \ref{fig:2}(d),
the key VBs and CBs are mainly contributed by the $X$-$p_{y,z}$ states of the distorted square nets.
Detailed calculations show that it is a nodal-line semimetal with two twisted nodal wires.
Each nodal wire consists of two noncontractible nodal lines traversing the bulk BZ, denoted by the red and blue lines in Fig. \ref{fig:2}(a).
With $\calT$ and $\calI$, the twofold degenerate nodal lines are protected by the combined antiunitary symmetry with $\pqty{\calT\calI}^2=1$.
Hereafter, we focus on the discussion of PrAsS in the main text.
More results of related materials are presented in Sec. B of the Supplemental Material (SM) \cite{supmat}.

\subsection{Symmetry analysis of fourfold degeneracy}
In PrAsS, the two noncontractible nodal lines cross each other at P $(\frac{\pi}{a}, k_y=0.0355\frac{2\pi}{b}, \frac{\pi}{c})$ on the U--R line [Fig. \ref{fig:2}(a)], leading to an unprecedented twisted nodal wire.
The computed irreducible representations (irreps) show that the crossing point P is accidentally fourfold degenerate,
formed by two two-dimensional-irrep bands (P1P2 and P3P4) with opposite $\tilde{C}_{2y}$ eigenvalues \cite{gao2021}.
The symmetries $\tilde{C}_{2y}$ and $\calT\tilde{C}_{2z}$ are preserved along the U--R line with
\begin{subequations}
\begin{align}
    \tilde{C}_{2y} \equiv & \; \{C_{2y}|0,1/2,0\}, \\
    \mathcal{T}\tilde{C}_{2z} \equiv & \; \{\mathcal{T}C_{2z}|1/2,0,1/2\}, \\
    \tilde{C}_{2y} \mathcal{T}\tilde{C}_{2z} = & \; \{E|010\} \mathcal{T} \tilde{C}_{2z} \tilde{C}_{2y}.
\end{align}
\end{subequations}
First, $\mathcal{T}\tilde{C}_{2z}$ will induce the Kramers-like degeneracy since $[\mathcal{T}\tilde{C}_{2z}]^{2}=-1$ on the whole U--R line.
Second, $\mathcal{T}\tilde{C}_{2z}$ related doublets share the same $\tilde{C}_{2y}$ eigenvalue. This can be deduced as the following.
With $(\tilde{C}_{2y})^2=\{E|010\}$, the eigenvalues of $\tilde{C}_{2y}$ are $\pm e^{-ik_y/2}$ at P.
Assuming wave function $\ket{\phi_P}$ has $\tilde{C}_{2y}$ eigenvalue $e^{-ik_y/2}$, then
\begin{equation}
    \begin{aligned}
        \tilde{C}_{2y}(\mathcal{T}\tilde{C}_{2z}\ket{\phi_P})
        = & \; \{E|010\}\mathcal{T}\tilde{C}_{2z}\tilde{C}_{2y}\ket{\phi_P} \\
        = & \; e^{-ik_y} (e^{-ik_y/2})^* (\mathcal{T}\tilde{C}_{2z}\ket{\phi_P})\\
        = & \; e^{-ik_y/2} (\mathcal{T}\tilde{C}_{2z}\ket{\phi_P}).
    \end{aligned}
\end{equation}
Thus, bands along U--R are always doubly degenerate with the $\mathcal{T}\tilde{C}_{2z}$-related doublets sharing the same $\tilde{C}_{2y}$ eigenvalue, $\{+,+\}e^{-i\frac{k_y}{2}}$ or $\{-,-\}e^{-i\frac{k_y}{2}}$.
Hence, the fourfold degeneracy at P comes from two $\mathcal{T}\tilde{C}_{2z}$ enforced doubly degenerate bands with the opposite $\tilde{C}_{2y}$ eigenvalues and is protected by both $\tilde{C}_{2y}$ and $\mathcal{T}\tilde{C}_{2z}$ symmetries.

\subsection{Nontrivial topology in the presence of SOC}
Upon including SOC, the two twisted nodal wires become fully gapped. The SIs ($z_2z_2z_2;z_4$) \cite{PhysRevB.76.045302FK, SymmIndicator2017Po, songzd2018natcomm, SymmIndicatorData2019Tang, SymmIndi2019ZhangTT} are then computed to be $(000;2)$, indicating a TCI phase \cite{Vergniory2019}.
Furthermore, MCNs $(m_0,m_\pi)$ can be defined in $k_{y}=0$ and $k_{y}=\pi/b$ planes with $M_y(\equiv \calI\tilde{C}_{2y})$ symmetry.
Using the Wilson-loop method, they are calculated to be $(m_{0},m_{\pi})=(0,0)$ (Figs. S5(a) and S5(b) in Sec. E of SM \cite{supmat}).
In addition, the hourglass invariant defined by the glide mirror operation $g_z(\equiv\calI \tilde{C}_{2z})$ is calculated to be 1 [Figs. \ref{fig:3}(a) and \ref{fig:3}(b)].
Therefore, we can expect the existence of the hourglass-shaped surface states \cite{Wang2016Hourglass, PhysRevX.6.021008, PhysRevB.94.155148, PhysRevB.101.115145} in (010)-surface BZ.

\subsection{QSH phase in the distorted $X$ square net}
The BS of a distorted square net ($xy$ plane) can be simply simulated by a $p_{x,y}$-based model with Slater-Koster \cite{PhysRev.94.1498LCAO} parameters.
There are two sites in a unit cell [$A:\pqty{0, 0}$ and $B: \pqty{1/2-\delta, 1/2}$ in Figs. \ref{fig:3}(c) and \ref{fig:1}(d)] with the parameter $\delta$ describing the distortion.
The nearest-neighbor (NN) bonds are given in Fig. \ref{fig:3}(c), forming zigzag chains, while the next-nearest-neighbor (NNN) bonds are indicated by dashed lines.
Each site contains $p_x$ and $p_y$ orbitals.
The hoppings for NN (NNN) bonds are given by the Slater-Koster parameters, $V^{-(+)}_{pp\sigma, pp\pi}$,
\begin{equation}
    \begin{aligned}
        V^{\mp}_{i} \equiv & \; \frac{l^2/2}{\pqty{l/2}^2 + \pqty{l/2\mp\delta ~l}^2}V_{i},\quad i\in\Bqty{pp\sigma, pp\pi}.
    \end{aligned}
\end{equation}
The onsite SOC term is given in the form of
\begin{equation}
    h_{\text{so}} = \lambda_{\text{so}}\: s_z \otimes \tau_0 \otimes \sigma_y,
\end{equation}
with $\bs$, $\btau$, and $\bsig$ Pauli matrices in spin, sub-lattice, and orbital space, respectively.
Thus, our distorted $p_{x,y}$ model are simply parametrized by $V_{pp\sigma, pp\pi}$, displacement $\delta$, and SOC strength $\lambda_{\text{so}}$.
The $V_{pp\sigma,pp\pi}$ parameters are extracted from first-principles calculations (Table S5 in Sec. F of the SM \cite{supmat}).
More details of the model can be found in Sec. C of the SM \cite{supmat}.
According to the phase diagram (\ie $|V_{pp\pi}/V_{pp\sigma}|\sim 0.3$ for Sb) in Fig. \ref{fig:1}(f), the QSH phase stands within a large area near the origin.
It shows that a small distortion $\delta$ (Jahn-Teller effect) enlarges the nontrivial gap for a given $\lambda_{\text{so}}$ (\eg $0.15\abs{V_{pp\sigma}}$).

\subsection{Minimum tight-binding model of the bulk}
The nontrivial band topology of $MXZ$ and $ABX_2$ compounds with and without SOC can be well reproduced by simply coupling two  distorted $X$ square-net layers, although these hoppings are very weak (see more details in Sec. D of the SM \cite{supmat}).
Its BS with SOC [Fig. \ref{fig:3}(d)] is similar to the $p_{y,z}$-fatted bands [Fig. \ref{fig:2}(c) and \ref{fig:2}(d)].
According to Refs. \cite{gao2021,Vergniory2019}, the SIs are computed to be $(000;2)$.
The $M_y$ MCNs and $g_z$ invariant are computed by the Wilson-loop method \cite{rspa.1984.0023, PhysRevLett.62.2747, PhysRevB.84.075119, PhysRevB.89.155114, PhysRevB.93.205104, PhysRevLett.107.036601}, which are identical with those of PrAsS from first-principles calculations.
As we expected, the hourglass-shaped surface states are obtained \cite{WU2017} in the (010)-surface spectrum [Fig. \ref{fig:3}(e)].

\subsection{3D QSH effect in $MXZ$ and $ABX_2$ compounds}
Since the interlayer coupling is weak due to the large distance $d$ between distorted $X$ square-net layers (\eg $\sim 9\AAA$ in PrAsS and $11\AAA$ in EuZnSb$_2$), one can simply consider the $MXZ$ and $ABX_2$ compounds as a stacking of QSH layers \cite{PhysRevB.91.081111, Qian2020}.
We notice that in real materials, there could be two difficulties in observing the 3D QSH effect experimentally.
First, the bulk states could be metallic, when the small QSH gap is messed out by other trivial bands.
However, there are a large family of these compounds sharing the same band topology (Fig. S1 and Table S1 in Sec. B of the SM \cite{supmat}), which allows us to adopt various chemical dopings.
In particular, crystals of CeAs$_{1+x}$Se$_{1-y}$ and EuCdSb$_2$ compounds with an insulating behavior have been synthesized successfully in $Pnma$ structures \cite{CeAsSe_Synthesis, Ohno_2021}.

\begin{figure}[!t]
    \centering
    \includegraphics[width=0.45\textwidth]{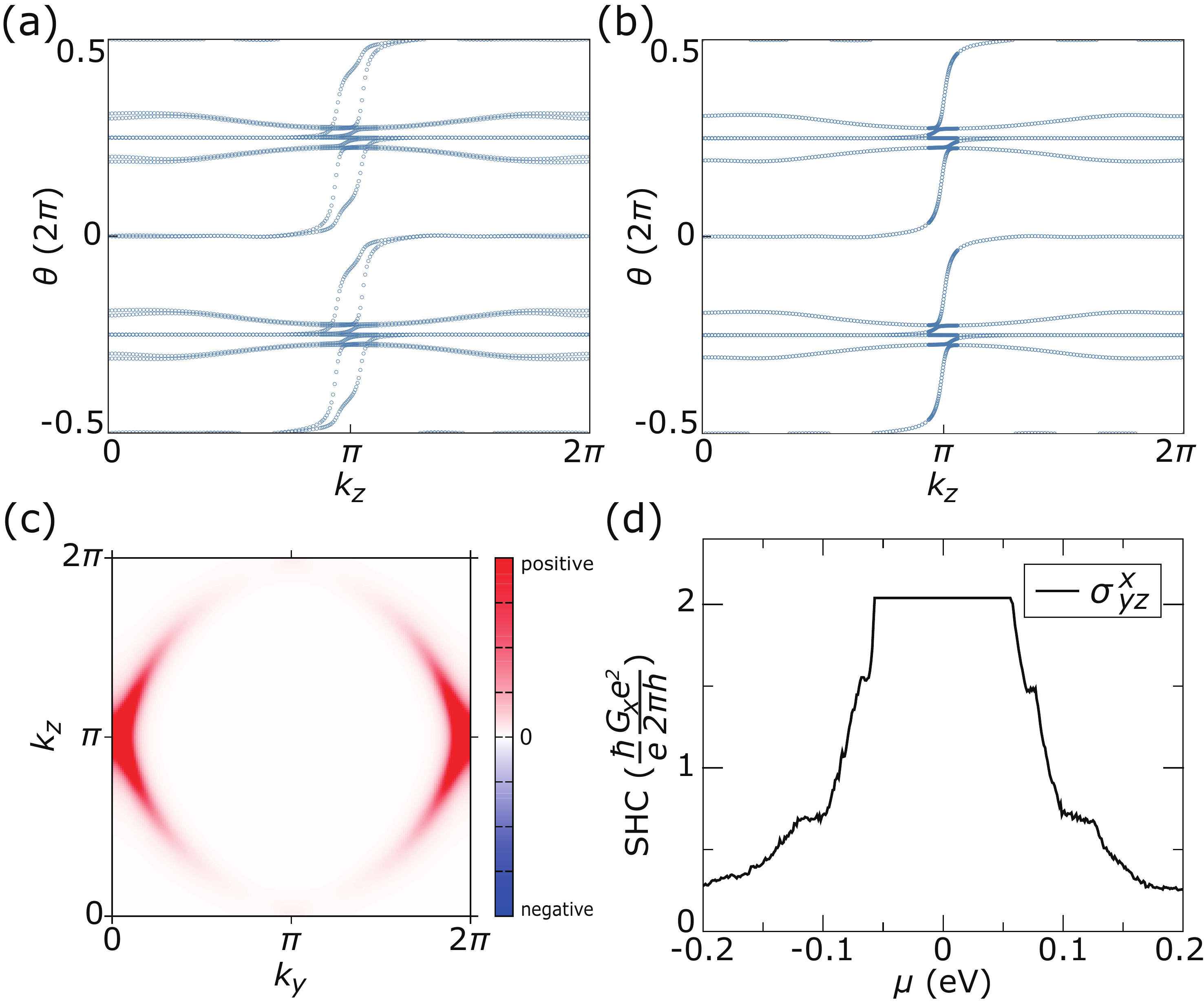}
    \caption{%
        (color online)
        The $k_y$-directed Wilson-loop spectrum in the positive-eigenvalue set of $\hat{s}_{x}$ for bulk PrAsS at
        (a) $k_{x} = 0$ and (b) $k_{x} = \pi/a$ planes, suggesting $C_{s+}=+2$.
        Note that the Wilson-loop bands in the $k_{x} = \pi/a$ plane host $\tilde{C}_{2x} \calT$-protected twofold degeneracy.
        (c) The distribution of $\sum_n f_{n}\pqty{\bk}\Omega^{x}_{yz;n}\pqty{\bk}$ in $k_x=\pi/a$ plane.
        (d) The calculated SHC $\sigma^x_{yz}$ as a function of chemical potential $\mu$.
        }
    \label{fig:4}
\end{figure}

Second, the $f$ electrons from $M$ or $A$ atoms can introduce bothersome magnetism to the systems.
However, the transition from a ($\calT$-broken) QSH phase to a trivial insulating state cannot happen without closing the band gap.
Since the $f$ electrons are quite localized and far below $E_F$ (\ie weakly coupled with the $X$-$p$ electrons), we believe that the $\calT$-broken QSH effect can be realized in the magnetism-weak-coupling limit \cite{PhysRevLett.107.066602}.
Here, we generalize the SCN defined in Ref. \cite{PhysRevLett.107.066602} to multiple-band systems.
As long as a spectrum gap exists between two sets of eigenvalues of $\hat{s}_{x}$ matrix presentation, the SCN ($C_{s+/-}$) is well defined for the positive/negative set.
The nontrivial topology can then be described by the generalized SCNs.
We further confirm $C_{s\pm}=\pm 2$ for each $k_x$-fixed plane using the Wilson-loop method.
The results of $k_x=0$ and $k_x=\pi/a$ planes are presented in Figs. \ref{fig:4}(a) and \ref{fig:4}(b).

We propose that such a 3D QSH effect can be realized in insulating crystals of these two families.
The TB Hamiltonian fully respects the symmetry and topology of corresponding materials, which is crucial to compute the intrinsic SHC.
Based on this TB model, we employed the Kubo formula approach at the clean limit to calculate the SHC \cite{zhang2017, RevModPhys.82.1959, PhysRevB.47.1651, PhysRevB.48.4442, berry_phase_David} of the TB model,
\begin{equation}
    \begin{aligned}
        \sigma_{\alpha\beta}^{\gamma} = & \frac{e}{\hbar} \sum_n \int_{\text{BZ}} \frac{\dd{\bk}}{\pqty{2\pi}^3} f_{n}\pqty{\bk} \Omega^{\gamma}_{\alpha\beta;n}\pqty{\bk}, \\
        \Omega^{\gamma}_{\alpha\beta;n}\pqty{\bk} = & 2i\hbar^2 \sum_{m\neq n} \frac{ \mel{u^n_{\bk}}{\hat{J}_{\alpha}^{\gamma}}{u^m_{\bk}} \mel{u^m_{\bk}}{\hat{v}_{\beta}}{u^n_{\bk}} }{ \pqty{\varepsilon^n_{\bk}-\varepsilon^m_{\bk}}^2 }
    \end{aligned}
\end{equation}
where $\hat{J}^\gamma_\alpha=\frac{1}{2}\Bqty{\hat{v}_\alpha,\hat{s}_\gamma}$ is the spin current operator, with $\hat{s}$ the spin operator, $\hat{v}_{\alpha} = \frac{\partial H}{\hbar \partial k_\alpha}$ the velocity operator, and $\alpha, \beta, \gamma=\Bqty{x,y,z}$.
$f_{n}\pqty{\bk}$ is the Fermi-Dirac distribution.
$\ket{u_{\bk}^n}$ and $\varepsilon^n_{\bk}$ are the eigenvectors and eigenvalues of Hamiltonian $h\pqty{\bk}$, respectively.
The distribution $\sum_n f_n\pqty{\bk}\Omega^x_{yz;n}\pqty{\bk}$ in $k_x=\pi/a$ plane is presented in Fig. \ref{fig:4}(c).
The SHC, exhibiting quantization, is calculated to be $\sigma^{x}_{yz} = (\frac{\hbar}{e}) \frac{2e^2}{h}\frac{G_x}{2\pi}$ with the chemical potential $\mu$ in the bulk gap [Fig. \ref{fig:4}(d)].
Here, $G_x$ is the $\hbx$ component of a reciprocal lattice vector.
As we know, the SHC is quantized only if the $\hat{s}_{x}$ is conserved.
First, in an $X$ layer, mirror-$x$ symmetry is slightly broken due to the weak buckled structure in the materials.
This symmetry can prohibit the intralayer hybridization between different spin channels in the basis of $p_{y,z}$ orbitals.
Second, the interlayer hoppings are very weak due to the large distance. Hence, we obtain a system nearly conserving $\hat{s}_{x}$, which yields the quantized SHC.

\section{Discussion}
Similarly, the distorted $X$ square nets can be also found in $I2mm$ structures (deviated from the $I4/mmm$ tetragonal structure), for instance, BaMnSb$_2$, where a 3D quantum Hall effect has been observed under magnetic fields \cite{Liu2021,pnas.1706657114, apl112111, PhysRevB.87.245104}.
On the other hand, the nontrivial distorted $X$ square net can be widely found in materials database, including superconductors, \eg the 112 family of iron pnictides Ca$_{1-x}$La$_x$FeAs$_2$ \cite{PhysRevB.91.081111, supr-con_112_jpn} and CaSb$_2$ \cite{supr-con_ax2_jpn,supr-con_ax2_2_jpn}.
The combination of nontrivial band topology and superconductivity may serve a platform for the search of intrinsic topological superconductivity and Majorana zero modes \cite{PhysRevX.10.041014, fese2015, tase2018, PhysrevX.12.011030}.

In summary, we find that the distorted $X$ square-net layers in $MXZ$ and $ABX_2$ compounds are QSH layers and the nontrivial topology relies on the $p_{y,z}$ orbitals of $X$ atoms.
These compounds can be simply considered as a stacking of QSH layers along the $\hbx$ direction.
Without SOC, the system hosts two twisted nodal wires, each of which contains two noncontractible nodal lines, crossing each other at a fourfold degenerate point.
Once SOC is taken into consideration, it becomes a TCI with SIs $(000;2)$ and MCNs $(0,0)$.
In the magnetism-weak-coupling limit, the nontrivial topology is characterized by the generalized SCNs $C_{s\pm}=\pm 2$.
The 3D QSH effect in these layered materials has been suggested by the calculated SHC,
which is promising in insulating compounds, like CeAs$_{1+x}$Se$_{1-y}$ and EuCdSb$_2$.

\begin{acknowledgments}
This work was supported by the National Natural Science Foundation of China (Grants No. 11974395 and No. 12188101), the Strategic Priority Research Program of Chinese Academy of Sciences (Grant No. XDB33000000), and the Center for Materials Genome.
C. Y. was supported by the Swiss National Science Foundation (SNF No. Grant 200021-196966).
H.W. acknowledges support from the Science Challenge Project (No. TZ2016004) and the K. C. Wong Education Foundation (No. GJTD-2018-01).
\end{acknowledgments}



\begin{thebibliography}{83}%
\makeatletter
\providecommand \@ifxundefined [1]{%
 \@ifx{#1\undefined}
}%
\providecommand \@ifnum [1]{%
 \ifnum #1\expandafter \@firstoftwo
 \else \expandafter \@secondoftwo
 \fi
}%
\providecommand \@ifx [1]{%
 \ifx #1\expandafter \@firstoftwo
 \else \expandafter \@secondoftwo
 \fi
}%
\providecommand \natexlab [1]{#1}%
\providecommand \enquote  [1]{``#1''}%
\providecommand \bibnamefont  [1]{#1}%
\providecommand \bibfnamefont [1]{#1}%
\providecommand \citenamefont [1]{#1}%
\providecommand \href@noop [0]{\@secondoftwo}%
\providecommand \href [0]{\begingroup \@sanitize@url \@href}%
\providecommand \@href[1]{\@@startlink{#1}\@@href}%
\providecommand \@@href[1]{\endgroup#1\@@endlink}%
\providecommand \@sanitize@url [0]{\catcode `\\12\catcode `\$12\catcode
  `\&12\catcode `\#12\catcode `\^12\catcode `\_12\catcode `\%12\relax}%
\providecommand \@@startlink[1]{}%
\providecommand \@@endlink[0]{}%
\providecommand \url  [0]{\begingroup\@sanitize@url \@url }%
\providecommand \@url [1]{\endgroup\@href {#1}{\urlprefix }}%
\providecommand \urlprefix  [0]{URL }%
\providecommand \Eprint [0]{\href }%
\providecommand \doibase [0]{http://dx.doi.org/}%
\providecommand \selectlanguage [0]{\@gobble}%
\providecommand \bibinfo  [0]{\@secondoftwo}%
\providecommand \bibfield  [0]{\@secondoftwo}%
\providecommand \translation [1]{[#1]}%
\providecommand \BibitemOpen [0]{}%
\providecommand \bibitemStop [0]{}%
\providecommand \bibitemNoStop [0]{.\EOS\space}%
\providecommand \EOS [0]{\spacefactor3000\relax}%
\providecommand \BibitemShut  [1]{\csname bibitem#1\endcsname}%
\let\auto@bib@innerbib\@empty
\bibitem [{\citenamefont {Hasan}\ and\ \citenamefont
  {Kane}(2010)}]{TIs-Hasan2010}%
  \BibitemOpen
  \bibfield  {author} {\bibinfo {author} {\bibfnamefont {M.~Z.}\ \bibnamefont
  {Hasan}}\ and\ \bibinfo {author} {\bibfnamefont {C.~L.}\ \bibnamefont
  {Kane}},\ }\href {\doibase 10.1103/RevModPhys.82.3045} {\bibfield  {journal}
  {\bibinfo  {journal} {Rev. Mod. Phys.}\ }\textbf {\bibinfo {volume} {82}},\
  \bibinfo {pages} {3045} (\bibinfo {year} {2010})}\BibitemShut {NoStop}%
\bibitem [{\citenamefont {Qi}\ and\ \citenamefont
  {Zhang}(2011)}]{TIs-FuLiang2011}%
  \BibitemOpen
  \bibfield  {author} {\bibinfo {author} {\bibfnamefont {X.-L.}\ \bibnamefont
  {Qi}}\ and\ \bibinfo {author} {\bibfnamefont {S.-C.}\ \bibnamefont {Zhang}},\
  }\href {\doibase 10.1103/RevModPhys.83.1057} {\bibfield  {journal} {\bibinfo
  {journal} {Rev. Mod. Phys.}\ }\textbf {\bibinfo {volume} {83}},\ \bibinfo
  {pages} {1057} (\bibinfo {year} {2011})}\BibitemShut {NoStop}%
\bibitem [{\citenamefont {Armitage}\ \emph {et~al.}(2018)\citenamefont
  {Armitage}, \citenamefont {Mele},\ and\ \citenamefont
  {Vishwanath}}]{Armitage2018}%
  \BibitemOpen
  \bibfield  {author} {\bibinfo {author} {\bibfnamefont {N.~P.}\ \bibnamefont
  {Armitage}}, \bibinfo {author} {\bibfnamefont {E.~J.}\ \bibnamefont {Mele}},
  \ and\ \bibinfo {author} {\bibfnamefont {A.}~\bibnamefont {Vishwanath}},\
  }\href {\doibase 10.1103/RevModPhys.90.015001} {\bibfield  {journal}
  {\bibinfo  {journal} {Rev. Mod. Phys.}\ }\textbf {\bibinfo {volume} {90}},\
  \bibinfo {pages} {015001} (\bibinfo {year} {2018})}\BibitemShut {NoStop}%
\bibitem [{\citenamefont {Bernevig}(2013)}]{BernevigTIandTSC}%
  \BibitemOpen
  \bibfield  {author} {\bibinfo {author} {\bibfnamefont {B.~A.}\ \bibnamefont
  {Bernevig}},\ }\href {\doibase doi:10.1515/9781400846733} {\emph {\bibinfo
  {title} {Topological Insulators and Topological Superconductors}}}\ (\bibinfo
   {publisher} {Princeton University Press, Princeton},\ \bibinfo {year}
  {2013})\BibitemShut {NoStop}%
\bibitem [{\citenamefont {Nie}\ \emph {et~al.}(2021{\natexlab{a}})\citenamefont
  {Nie}, \citenamefont {Bernevig},\ and\ \citenamefont
  {Wang}}]{PhysRevResearch.3.L012028}%
  \BibitemOpen
  \bibfield  {author} {\bibinfo {author} {\bibfnamefont {S.}~\bibnamefont
  {Nie}}, \bibinfo {author} {\bibfnamefont {B.~A.}\ \bibnamefont {Bernevig}}, \
  and\ \bibinfo {author} {\bibfnamefont {Z.}~\bibnamefont {Wang}},\ }\href
  {\doibase 10.1103/PhysRevResearch.3.L012028} {\bibfield  {journal} {\bibinfo
  {journal} {Phys. Rev. Research}\ }\textbf {\bibinfo {volume} {3}},\ \bibinfo
  {pages} {L012028} (\bibinfo {year} {2021}{\natexlab{a}})}\BibitemShut
  {NoStop}%
\bibitem [{\citenamefont {Guo}\ \emph {et~al.}(2021)\citenamefont {Guo},
  \citenamefont {Yan}, \citenamefont {Sheng}, \citenamefont {Nie},
  \citenamefont {Shi},\ and\ \citenamefont {Wang}}]{PhysRevB.103.115145}%
  \BibitemOpen
  \bibfield  {author} {\bibinfo {author} {\bibfnamefont {Z.}~\bibnamefont
  {Guo}}, \bibinfo {author} {\bibfnamefont {D.}~\bibnamefont {Yan}}, \bibinfo
  {author} {\bibfnamefont {H.}~\bibnamefont {Sheng}}, \bibinfo {author}
  {\bibfnamefont {S.}~\bibnamefont {Nie}}, \bibinfo {author} {\bibfnamefont
  {Y.}~\bibnamefont {Shi}}, \ and\ \bibinfo {author} {\bibfnamefont
  {Z.}~\bibnamefont {Wang}},\ }\href {\doibase 10.1103/PhysRevB.103.115145}
  {\bibfield  {journal} {\bibinfo  {journal} {Phys. Rev. B}\ }\textbf {\bibinfo
  {volume} {103}},\ \bibinfo {pages} {115145} (\bibinfo {year}
  {2021})}\BibitemShut {NoStop}%
\bibitem [{\citenamefont {Qian}\ \emph
  {et~al.}(2020{\natexlab{a}})\citenamefont {Qian}, \citenamefont {Gao},
  \citenamefont {Song}, \citenamefont {Nie}, \citenamefont {Wang},
  \citenamefont {Weng},\ and\ \citenamefont {Fang}}]{PhysRevB.101.155143}%
  \BibitemOpen
  \bibfield  {author} {\bibinfo {author} {\bibfnamefont {Y.}~\bibnamefont
  {Qian}}, \bibinfo {author} {\bibfnamefont {J.}~\bibnamefont {Gao}}, \bibinfo
  {author} {\bibfnamefont {Z.}~\bibnamefont {Song}}, \bibinfo {author}
  {\bibfnamefont {S.}~\bibnamefont {Nie}}, \bibinfo {author} {\bibfnamefont
  {Z.}~\bibnamefont {Wang}}, \bibinfo {author} {\bibfnamefont {H.}~\bibnamefont
  {Weng}}, \ and\ \bibinfo {author} {\bibfnamefont {Z.}~\bibnamefont {Fang}},\
  }\href {\doibase 10.1103/PhysRevB.101.155143} {\bibfield  {journal} {\bibinfo
   {journal} {Phys. Rev. B}\ }\textbf {\bibinfo {volume} {101}},\ \bibinfo
  {pages} {155143} (\bibinfo {year} {2020}{\natexlab{a}})}\BibitemShut
  {NoStop}%
\bibitem [{\citenamefont {Gao}\ \emph {et~al.}(2021{\natexlab{a}})\citenamefont
  {Gao}, \citenamefont {Qian}, \citenamefont {Nie}, \citenamefont {Fang},
  \citenamefont {Weng},\ and\ \citenamefont {Wang}}]{GAO2021667}%
  \BibitemOpen
  \bibfield  {author} {\bibinfo {author} {\bibfnamefont {J.}~\bibnamefont
  {Gao}}, \bibinfo {author} {\bibfnamefont {Y.}~\bibnamefont {Qian}}, \bibinfo
  {author} {\bibfnamefont {S.}~\bibnamefont {Nie}}, \bibinfo {author}
  {\bibfnamefont {Z.}~\bibnamefont {Fang}}, \bibinfo {author} {\bibfnamefont
  {H.}~\bibnamefont {Weng}}, \ and\ \bibinfo {author} {\bibfnamefont
  {Z.}~\bibnamefont {Wang}},\ }\href {\doibase
  https://doi.org/10.1016/j.scib.2020.12.028} {\bibfield  {journal} {\bibinfo
  {journal} {Sci. Bull.}\ }\textbf {\bibinfo {volume} {66}},\ \bibinfo
  {pages} {667} (\bibinfo {year} {2021}{\natexlab{a}})}\BibitemShut {NoStop}%
\bibitem [{\citenamefont {Bradlyn}\ \emph {et~al.}(2017)\citenamefont
  {Bradlyn}, \citenamefont {Elcoro}, \citenamefont {Cano}, \citenamefont
  {Vergniory}, \citenamefont {Wang}, \citenamefont {Felser}, \citenamefont
  {Aroyo},\ and\ \citenamefont {Bernevig}}]{Bradlyn2017}%
  \BibitemOpen
  \bibfield  {author} {\bibinfo {author} {\bibfnamefont {B.}~\bibnamefont
  {Bradlyn}}, \bibinfo {author} {\bibfnamefont {L.}~\bibnamefont {Elcoro}},
  \bibinfo {author} {\bibfnamefont {J.}~\bibnamefont {Cano}}, \bibinfo {author}
  {\bibfnamefont {M.~G.}\ \bibnamefont {Vergniory}}, \bibinfo {author}
  {\bibfnamefont {Z.}~\bibnamefont {Wang}}, \bibinfo {author} {\bibfnamefont
  {C.}~\bibnamefont {Felser}}, \bibinfo {author} {\bibfnamefont {M.~I.}\
  \bibnamefont {Aroyo}}, \ and\ \bibinfo {author} {\bibfnamefont {B.~A.}\
  \bibnamefont {Bernevig}},\ }\href {\doibase 10.1038/nature23268} {\bibfield
  {journal} {\bibinfo  {journal} {Nature}\ }\textbf {\bibinfo {volume} {547}},\
  \bibinfo {pages} {298} (\bibinfo {year} {2017})}\BibitemShut {NoStop}%
\bibitem [{\citenamefont {Cano}\ \emph
  {et~al.}(2018{\natexlab{a}})\citenamefont {Cano}, \citenamefont {Bradlyn},
  \citenamefont {Wang}, \citenamefont {Elcoro}, \citenamefont {Vergniory},
  \citenamefont {Felser}, \citenamefont {Aroyo},\ and\ \citenamefont
  {Bernevig}}]{Bernevig-EBR-2018}%
  \BibitemOpen
  \bibfield  {author} {\bibinfo {author} {\bibfnamefont {J.}~\bibnamefont
  {Cano}}, \bibinfo {author} {\bibfnamefont {B.}~\bibnamefont {Bradlyn}},
  \bibinfo {author} {\bibfnamefont {Z.}~\bibnamefont {Wang}}, \bibinfo {author}
  {\bibfnamefont {L.}~\bibnamefont {Elcoro}}, \bibinfo {author} {\bibfnamefont
  {M.~G.}\ \bibnamefont {Vergniory}}, \bibinfo {author} {\bibfnamefont
  {C.}~\bibnamefont {Felser}}, \bibinfo {author} {\bibfnamefont {M.~I.}\
  \bibnamefont {Aroyo}}, \ and\ \bibinfo {author} {\bibfnamefont {B.~A.}\
  \bibnamefont {Bernevig}},\ }\href {\doibase 10.1103/PhysRevB.97.035139}
  {\bibfield  {journal} {\bibinfo  {journal} {Phys. Rev. B}\ }\textbf {\bibinfo
  {volume} {97}},\ \bibinfo {pages} {035139} (\bibinfo {year}
  {2018}{\natexlab{a}})}\BibitemShut {NoStop}%
\bibitem [{\citenamefont {Bradlyn}\ \emph {et~al.}(2018)\citenamefont
  {Bradlyn}, \citenamefont {Elcoro}, \citenamefont {Vergniory}, \citenamefont
  {Cano}, \citenamefont {Wang}, \citenamefont {Felser}, \citenamefont {Aroyo},\
  and\ \citenamefont {Bernevig}}]{Bernevig-band-graphtheory-2018}%
  \BibitemOpen
  \bibfield  {author} {\bibinfo {author} {\bibfnamefont {B.}~\bibnamefont
  {Bradlyn}}, \bibinfo {author} {\bibfnamefont {L.}~\bibnamefont {Elcoro}},
  \bibinfo {author} {\bibfnamefont {M.~G.}\ \bibnamefont {Vergniory}}, \bibinfo
  {author} {\bibfnamefont {J.}~\bibnamefont {Cano}}, \bibinfo {author}
  {\bibfnamefont {Z.}~\bibnamefont {Wang}}, \bibinfo {author} {\bibfnamefont
  {C.}~\bibnamefont {Felser}}, \bibinfo {author} {\bibfnamefont {M.~I.}\
  \bibnamefont {Aroyo}}, \ and\ \bibinfo {author} {\bibfnamefont {B.~A.}\
  \bibnamefont {Bernevig}},\ }\href {\doibase 10.1103/PhysRevB.97.035138}
  {\bibfield  {journal} {\bibinfo  {journal} {Phys. Rev. B}\ }\textbf {\bibinfo
  {volume} {97}},\ \bibinfo {pages} {035138} (\bibinfo {year}
  {2018})}\BibitemShut {NoStop}%
\bibitem [{\citenamefont {Nie}\ \emph {et~al.}(2021{\natexlab{b}})\citenamefont
  {Nie}, \citenamefont {Qian}, \citenamefont {Gao}, \citenamefont {Fang},
  \citenamefont {Weng},\ and\ \citenamefont {Wang}}]{eletride2021}%
  \BibitemOpen
  \bibfield  {author} {\bibinfo {author} {\bibfnamefont {S.}~\bibnamefont
  {Nie}}, \bibinfo {author} {\bibfnamefont {Y.}~\bibnamefont {Qian}}, \bibinfo
  {author} {\bibfnamefont {J.}~\bibnamefont {Gao}}, \bibinfo {author}
  {\bibfnamefont {Z.}~\bibnamefont {Fang}}, \bibinfo {author} {\bibfnamefont
  {H.}~\bibnamefont {Weng}}, \ and\ \bibinfo {author} {\bibfnamefont
  {Z.}~\bibnamefont {Wang}},\ }\href {\doibase 10.1103/PhysRevB.103.205133}
  {\bibfield  {journal} {\bibinfo  {journal} {Phys. Rev. B}\ }\textbf {\bibinfo
  {volume} {103}},\ \bibinfo {pages} {205133} (\bibinfo {year}
  {2021}{\natexlab{b}})}\BibitemShut {NoStop}%
\bibitem [{\citenamefont {Bernevig}\ \emph {et~al.}(2006)\citenamefont
  {Bernevig}, \citenamefont {Hughes},\ and\ \citenamefont
  {Zhang}}]{BHZScience}%
  \BibitemOpen
  \bibfield  {author} {\bibinfo {author} {\bibfnamefont {B.~A.}\ \bibnamefont
  {Bernevig}}, \bibinfo {author} {\bibfnamefont {T.~L.}\ \bibnamefont
  {Hughes}}, \ and\ \bibinfo {author} {\bibfnamefont {S.-C.}\ \bibnamefont
  {Zhang}},\ }\href {\doibase 10.1126/science.1133734} {\bibfield  {journal}
  {\bibinfo  {journal} {Science}\ }\textbf {\bibinfo {volume} {314}},\ \bibinfo
  {pages} {1757} (\bibinfo {year} {2006})}
  \BibitemShut
  {NoStop}%
\bibitem [{\citenamefont {K{\"o}nig}\ \emph {et~al.}(2007)\citenamefont
  {K{\"o}nig}, \citenamefont {Wiedmann}, \citenamefont {Br{\"u}ne},
  \citenamefont {Roth}, \citenamefont {Buhmann}, \citenamefont {Molenkamp},
  \citenamefont {Qi},\ and\ \citenamefont {Zhang}}]{BHZScienceExp}%
  \BibitemOpen
  \bibfield  {author} {\bibinfo {author} {\bibfnamefont {M.}~\bibnamefont
  {K{\"o}nig}}, \bibinfo {author} {\bibfnamefont {S.}~\bibnamefont {Wiedmann}},
  \bibinfo {author} {\bibfnamefont {C.}~\bibnamefont {Br{\"u}ne}}, \bibinfo
  {author} {\bibfnamefont {A.}~\bibnamefont {Roth}}, \bibinfo {author}
  {\bibfnamefont {H.}~\bibnamefont {Buhmann}}, \bibinfo {author} {\bibfnamefont
  {L.~W.}\ \bibnamefont {Molenkamp}}, \bibinfo {author} {\bibfnamefont {X.-L.}\
  \bibnamefont {Qi}}, \ and\ \bibinfo {author} {\bibfnamefont {S.-C.}\
  \bibnamefont {Zhang}},\ }\href {\doibase 10.1126/science.1148047} {\bibfield
  {journal} {\bibinfo  {journal} {Science}\ }\textbf {\bibinfo {volume}
  {318}},\ \bibinfo {pages} {766} (\bibinfo {year} {2007})}
  \BibitemShut
  {NoStop}%
\bibitem [{\citenamefont {Novik}\ \emph {et~al.}(2005)\citenamefont {Novik},
  \citenamefont {Pfeuffer-Jeschke}, \citenamefont {Jungwirth}, \citenamefont
  {Latussek}, \citenamefont {Becker}, \citenamefont {Landwehr}, \citenamefont
  {Buhmann},\ and\ \citenamefont {Molenkamp}}]{PhysRevB.72.035321BHZpara}%
  \BibitemOpen
  \bibfield  {author} {\bibinfo {author} {\bibfnamefont {E.~G.}\ \bibnamefont
  {Novik}}, \bibinfo {author} {\bibfnamefont {A.}~\bibnamefont
  {Pfeuffer-Jeschke}}, \bibinfo {author} {\bibfnamefont {T.}~\bibnamefont
  {Jungwirth}}, \bibinfo {author} {\bibfnamefont {V.}~\bibnamefont {Latussek}},
  \bibinfo {author} {\bibfnamefont {C.~R.}\ \bibnamefont {Becker}}, \bibinfo
  {author} {\bibfnamefont {G.}~\bibnamefont {Landwehr}}, \bibinfo {author}
  {\bibfnamefont {H.}~\bibnamefont {Buhmann}}, \ and\ \bibinfo {author}
  {\bibfnamefont {L.~W.}\ \bibnamefont {Molenkamp}},\ }\href {\doibase
  10.1103/PhysRevB.72.035321} {\bibfield  {journal} {\bibinfo  {journal} {Phys.
  Rev. B}\ }\textbf {\bibinfo {volume} {72}},\ \bibinfo {pages} {035321}
  (\bibinfo {year} {2005})}\BibitemShut {NoStop}%
\bibitem [{\citenamefont {Zhang}\ \emph {et~al.}(2009)\citenamefont {Zhang},
  \citenamefont {Liu}, \citenamefont {Qi}, \citenamefont {Dai}, \citenamefont
  {Fang},\ and\ \citenamefont {Zhang}}]{Bi2X3-zhj2009}%
  \BibitemOpen
  \bibfield  {author} {\bibinfo {author} {\bibfnamefont {H.}~\bibnamefont
  {Zhang}}, \bibinfo {author} {\bibfnamefont {C.~X.}\ \bibnamefont {Liu}},
  \bibinfo {author} {\bibfnamefont {X.~L.}\ \bibnamefont {Qi}}, \bibinfo
  {author} {\bibfnamefont {X.}~\bibnamefont {Dai}}, \bibinfo {author}
  {\bibfnamefont {Z.}~\bibnamefont {Fang}}, \ and\ \bibinfo {author}
  {\bibfnamefont {S.~C.}\ \bibnamefont {Zhang}},\ }\href {\doibase
  10.1038/nphys1270} {\bibfield  {journal} {\bibinfo  {journal} {Nat. Phys.}\
  }\textbf {\bibinfo {volume} {5}},\ \bibinfo {pages} {438} (\bibinfo {year}
  {2009})}\BibitemShut {NoStop}%
\bibitem [{\citenamefont {Hsieh}\ \emph {et~al.}(2009)\citenamefont {Hsieh},
  \citenamefont {Xia}, \citenamefont {Qian}, \citenamefont {Wray},
  \citenamefont {Dil}, \citenamefont {Meier}, \citenamefont {Osterwalder},
  \citenamefont {Patthey}, \citenamefont {Checkelsky},\ and\ \citenamefont
  {Ong}}]{Hsieh2009}%
  \BibitemOpen
  \bibfield  {author} {\bibinfo {author} {\bibfnamefont {D.}~\bibnamefont
  {Hsieh}}, \bibinfo {author} {\bibfnamefont {Y.}~\bibnamefont {Xia}}, \bibinfo
  {author} {\bibfnamefont {D.}~\bibnamefont {Qian}}, \bibinfo {author}
  {\bibfnamefont {L.}~\bibnamefont {Wray}}, \bibinfo {author} {\bibfnamefont
  {J.~H.}\ \bibnamefont {Dil}}, \bibinfo {author} {\bibfnamefont
  {F.}~\bibnamefont {Meier}}, \bibinfo {author} {\bibfnamefont
  {J.}~\bibnamefont {Osterwalder}}, \bibinfo {author} {\bibfnamefont
  {L.}~\bibnamefont {Patthey}}, \bibinfo {author} {\bibfnamefont {J.~G.}\
  \bibnamefont {Checkelsky}}, \ and\ \bibinfo {author} {\bibfnamefont {N.~P.}\
  \bibnamefont {Ong}},\ }\href {\doibase 10.1038/nature08234} {\bibfield
  {journal} {\bibinfo  {journal} {Nature}\ }\textbf {\bibinfo {volume} {460}},\
  \bibinfo {pages} {1101} (\bibinfo {year} {2009})}\BibitemShut {NoStop}%
\bibitem [{\citenamefont {Xia}\ \emph {et~al.}(2009)\citenamefont {Xia},
  \citenamefont {Qian}, \citenamefont {Hsieh}, \citenamefont {Wray},
  \citenamefont {Pal}, \citenamefont {Lin}, \citenamefont {Bansil},
  \citenamefont {Grauer}, \citenamefont {Hor}, \citenamefont {Cava},\ and\
  \citenamefont {Hasan}}]{Bi2X3-Hasan2009}%
  \BibitemOpen
  \bibfield  {author} {\bibinfo {author} {\bibfnamefont {Y.}~\bibnamefont
  {Xia}}, \bibinfo {author} {\bibfnamefont {D.}~\bibnamefont {Qian}}, \bibinfo
  {author} {\bibfnamefont {D.}~\bibnamefont {Hsieh}}, \bibinfo {author}
  {\bibfnamefont {L.}~\bibnamefont {Wray}}, \bibinfo {author} {\bibfnamefont
  {A.}~\bibnamefont {Pal}}, \bibinfo {author} {\bibfnamefont {H.}~\bibnamefont
  {Lin}}, \bibinfo {author} {\bibfnamefont {A.}~\bibnamefont {Bansil}},
  \bibinfo {author} {\bibfnamefont {D.}~\bibnamefont {Grauer}}, \bibinfo
  {author} {\bibfnamefont {Y.~S.}\ \bibnamefont {Hor}}, \bibinfo {author}
  {\bibfnamefont {R.~J.}\ \bibnamefont {Cava}}, \ and\ \bibinfo {author}
  {\bibfnamefont {M.~Z.}\ \bibnamefont {Hasan}},\ }\href {\doibase
  10.1038/nphys1274} {\bibfield  {journal} {\bibinfo  {journal} {Nat. Phys.}\
  }\textbf {\bibinfo {volume} {5}},\ \bibinfo {pages} {398} (\bibinfo {year}
  {2009})}\BibitemShut {NoStop}%
\bibitem [{\citenamefont {Chen}\ \emph {et~al.}(2009)\citenamefont {Chen},
  \citenamefont {Analytis}, \citenamefont {Chu}, \citenamefont {Liu},
  \citenamefont {Mo}, \citenamefont {Qi}, \citenamefont {Zhang}, \citenamefont
  {Lu}, \citenamefont {Dai}, \citenamefont {Fang}, \citenamefont {Zhang},
  \citenamefont {Fisher}, \citenamefont {Hussain},\ and\ \citenamefont
  {Shen}}]{Bi2Te3-Shen2009}%
  \BibitemOpen
  \bibfield  {author} {\bibinfo {author} {\bibfnamefont {Y.~L.}\ \bibnamefont
  {Chen}}, \bibinfo {author} {\bibfnamefont {J.~G.}\ \bibnamefont {Analytis}},
  \bibinfo {author} {\bibfnamefont {J.-H.}\ \bibnamefont {Chu}}, \bibinfo
  {author} {\bibfnamefont {Z.~K.}\ \bibnamefont {Liu}}, \bibinfo {author}
  {\bibfnamefont {S.-K.}\ \bibnamefont {Mo}}, \bibinfo {author} {\bibfnamefont
  {X.~L.}\ \bibnamefont {Qi}}, \bibinfo {author} {\bibfnamefont {H.~J.}\
  \bibnamefont {Zhang}}, \bibinfo {author} {\bibfnamefont {D.~H.}\ \bibnamefont
  {Lu}}, \bibinfo {author} {\bibfnamefont {X.}~\bibnamefont {Dai}}, \bibinfo
  {author} {\bibfnamefont {Z.}~\bibnamefont {Fang}}, \bibinfo {author}
  {\bibfnamefont {S.~C.}\ \bibnamefont {Zhang}}, \bibinfo {author}
  {\bibfnamefont {I.~R.}\ \bibnamefont {Fisher}}, \bibinfo {author}
  {\bibfnamefont {Z.}~\bibnamefont {Hussain}}, \ and\ \bibinfo {author}
  {\bibfnamefont {Z.-X.}\ \bibnamefont {Shen}},\ }\href {\doibase
  10.1126/science.1173034} {\bibfield  {journal} {\bibinfo  {journal}
  {Science}\ }\textbf {\bibinfo {volume} {325}},\ \bibinfo {pages} {178}
  (\bibinfo {year} {2009})}\BibitemShut {NoStop}%
\bibitem [{\citenamefont {Kane}\ and\ \citenamefont
  {Mele}(2005{\natexlab{a}})}]{PhysRevLett.95.226801KaneMeleRef1}%
  \BibitemOpen
  \bibfield  {author} {\bibinfo {author} {\bibfnamefont {C.~L.}\ \bibnamefont
  {Kane}}\ and\ \bibinfo {author} {\bibfnamefont {E.~J.}\ \bibnamefont
  {Mele}},\ }\href {\doibase 10.1103/PhysRevLett.95.226801} {\bibfield
  {journal} {\bibinfo  {journal} {Phys. Rev. Lett.}\ }\textbf {\bibinfo
  {volume} {95}},\ \bibinfo {pages} {226801} (\bibinfo {year}
  {2005}{\natexlab{a}})}\BibitemShut {NoStop}%
\bibitem [{\citenamefont {Kane}\ and\ \citenamefont
  {Mele}(2005{\natexlab{b}})}]{PhysRevLett.95.146802KaneMeleRef2}%
  \BibitemOpen
  \bibfield  {author} {\bibinfo {author} {\bibfnamefont {C.~L.}\ \bibnamefont
  {Kane}}\ and\ \bibinfo {author} {\bibfnamefont {E.~J.}\ \bibnamefont
  {Mele}},\ }\href {\doibase 10.1103/PhysRevLett.95.146802} {\bibfield
  {journal} {\bibinfo  {journal} {Phys. Rev. Lett.}\ }\textbf {\bibinfo
  {volume} {95}},\ \bibinfo {pages} {146802} (\bibinfo {year}
  {2005}{\natexlab{b}})}\BibitemShut {NoStop}%
\bibitem [{\citenamefont {Liu}\ \emph {et~al.}(2011)\citenamefont {Liu},
  \citenamefont {Feng},\ and\ \citenamefont {Yao}}]{PhysRevLett.107.076802}%
  \BibitemOpen
  \bibfield  {author} {\bibinfo {author} {\bibfnamefont {C.-C.}\ \bibnamefont
  {Liu}}, \bibinfo {author} {\bibfnamefont {W.}~\bibnamefont {Feng}}, \ and\
  \bibinfo {author} {\bibfnamefont {Y.}~\bibnamefont {Yao}},\ }\href {\doibase
  10.1103/PhysRevLett.107.076802} {\bibfield  {journal} {\bibinfo  {journal}
  {Phys. Rev. Lett.}\ }\textbf {\bibinfo {volume} {107}},\ \bibinfo {pages}
  {076802} (\bibinfo {year} {2011})}\BibitemShut {NoStop}%
\bibitem [{\citenamefont {Zhou}\ \emph {et~al.}(2014)\citenamefont {Zhou},
  \citenamefont {Ming}, \citenamefont {Liu}, \citenamefont {Wang},
  \citenamefont {Li},\ and\ \citenamefont {Liu}}]{PNAS.140970111SiCRef1}%
  \BibitemOpen
  \bibfield  {author} {\bibinfo {author} {\bibfnamefont {M.}~\bibnamefont
  {Zhou}}, \bibinfo {author} {\bibfnamefont {W.}~\bibnamefont {Ming}}, \bibinfo
  {author} {\bibfnamefont {Z.}~\bibnamefont {Liu}}, \bibinfo {author}
  {\bibfnamefont {Z.}~\bibnamefont {Wang}}, \bibinfo {author} {\bibfnamefont
  {P.}~\bibnamefont {Li}}, \ and\ \bibinfo {author} {\bibfnamefont
  {F.}~\bibnamefont {Liu}},\ }\href {\doibase 10.1073/pnas.1409701111}
  {\bibfield  {journal} {\bibinfo  {journal} {Proc. Nat.
  Acad. Sci. USA}\ }\textbf {\bibinfo
  {volume} {111}} (\bibinfo {year} {2014})}
  \BibitemShut {NoStop}%
\bibitem [{\citenamefont {Hsu}\ \emph {et~al.}(2015)\citenamefont {Hsu},
  \citenamefont {Huang}, \citenamefont {Chuang}, \citenamefont {Kuo},
  \citenamefont {Liu}, \citenamefont {Lin},\ and\ \citenamefont
  {Bansil}}]{Hsu_2015SiCRef2}%
  \BibitemOpen
  \bibfield  {author} {\bibinfo {author} {\bibfnamefont {C.-H.}\ \bibnamefont
  {Hsu}}, \bibinfo {author} {\bibfnamefont {Z.-Q.}\ \bibnamefont {Huang}},
  \bibinfo {author} {\bibfnamefont {F.-C.}\ \bibnamefont {Chuang}}, \bibinfo
  {author} {\bibfnamefont {C.-C.}\ \bibnamefont {Kuo}}, \bibinfo {author}
  {\bibfnamefont {Y.-T.}\ \bibnamefont {Liu}}, \bibinfo {author} {\bibfnamefont
  {H.}~\bibnamefont {Lin}}, \ and\ \bibinfo {author} {\bibfnamefont
  {A.}~\bibnamefont {Bansil}},\ }\href {\doibase 10.1088/1367-2630/17/2/025005}
  {\bibfield  {journal} {\bibinfo  {journal} {New Journal of Physics}\ }\textbf
  {\bibinfo {volume} {17}},\ \bibinfo {pages} {025005} (\bibinfo {year}
  {2015})}\BibitemShut {NoStop}%
\bibitem [{\citenamefont {Reis}\ \emph {et~al.}(2017)\citenamefont {Reis},
  \citenamefont {Li}, \citenamefont {Dudy}, \citenamefont {Bauernfeind},
  \citenamefont {Glass}, \citenamefont {Hanke}, \citenamefont {Thomale},
  \citenamefont {Sch{\"a}fer},\ and\ \citenamefont {Claessen}}]{Reis287}%
  \BibitemOpen
  \bibfield  {author} {\bibinfo {author} {\bibfnamefont {F.}~\bibnamefont
  {Reis}}, \bibinfo {author} {\bibfnamefont {G.}~\bibnamefont {Li}}, \bibinfo
  {author} {\bibfnamefont {L.}~\bibnamefont {Dudy}}, \bibinfo {author}
  {\bibfnamefont {M.}~\bibnamefont {Bauernfeind}}, \bibinfo {author}
  {\bibfnamefont {S.}~\bibnamefont {Glass}}, \bibinfo {author} {\bibfnamefont
  {W.}~\bibnamefont {Hanke}}, \bibinfo {author} {\bibfnamefont
  {R.}~\bibnamefont {Thomale}}, \bibinfo {author} {\bibfnamefont
  {J.}~\bibnamefont {Sch{\"a}fer}}, \ and\ \bibinfo {author} {\bibfnamefont
  {R.}~\bibnamefont {Claessen}},\ }\href {\doibase 10.1126/science.aai8142}
  {\bibfield  {journal} {\bibinfo  {journal} {Science}\ }\textbf {\bibinfo
  {volume} {357}},\ \bibinfo {pages} {287} (\bibinfo {year}
  {2017})}\BibitemShut {NoStop}%
\bibitem [{\citenamefont {Bergman}\ \emph {et~al.}(2008)\citenamefont
  {Bergman}, \citenamefont {Wu},\ and\ \citenamefont
  {Balents}}]{PhysRevB.78.125104FlatBandTPTRef1}%
  \BibitemOpen
  \bibfield  {author} {\bibinfo {author} {\bibfnamefont {D.~L.}\ \bibnamefont
  {Bergman}}, \bibinfo {author} {\bibfnamefont {C.}~\bibnamefont {Wu}}, \ and\
  \bibinfo {author} {\bibfnamefont {L.}~\bibnamefont {Balents}},\ }\href
  {\doibase 10.1103/PhysRevB.78.125104} {\bibfield  {journal} {\bibinfo
  {journal} {Phys. Rev. B}\ }\textbf {\bibinfo {volume} {78}},\ \bibinfo
  {pages} {125104} (\bibinfo {year} {2008})}\BibitemShut {NoStop}%
\bibitem [{\citenamefont {Ma}\ \emph {et~al.}(2020)\citenamefont {Ma},
  \citenamefont {Rhim}, \citenamefont {Tang}, \citenamefont {Xia},
  \citenamefont {Wang}, \citenamefont {Zheng}, \citenamefont {Xia},
  \citenamefont {Song}, \citenamefont {Hu}, \citenamefont {Li}, \citenamefont
  {Yang}, \citenamefont {Leykam},\ and\ \citenamefont
  {Chen}}]{PhysRevLett.124.183901FlatBandTPTRef2}%
  \BibitemOpen
  \bibfield  {author} {\bibinfo {author} {\bibfnamefont {J.}~\bibnamefont
  {Ma}}, \bibinfo {author} {\bibfnamefont {J.-W.}\ \bibnamefont {Rhim}},
  \bibinfo {author} {\bibfnamefont {L.}~\bibnamefont {Tang}}, \bibinfo {author}
  {\bibfnamefont {S.}~\bibnamefont {Xia}}, \bibinfo {author} {\bibfnamefont
  {H.}~\bibnamefont {Wang}}, \bibinfo {author} {\bibfnamefont {X.}~\bibnamefont
  {Zheng}}, \bibinfo {author} {\bibfnamefont {S.}~\bibnamefont {Xia}}, \bibinfo
  {author} {\bibfnamefont {D.}~\bibnamefont {Song}}, \bibinfo {author}
  {\bibfnamefont {Y.}~\bibnamefont {Hu}}, \bibinfo {author} {\bibfnamefont
  {Y.}~\bibnamefont {Li}}, \bibinfo {author} {\bibfnamefont {B.-J.}\
  \bibnamefont {Yang}}, \bibinfo {author} {\bibfnamefont {D.}~\bibnamefont
  {Leykam}}, \ and\ \bibinfo {author} {\bibfnamefont {Z.}~\bibnamefont
  {Chen}},\ }\href {\doibase 10.1103/PhysRevLett.124.183901} {\bibfield
  {journal} {\bibinfo  {journal} {Phys. Rev. Lett.}\ }\textbf {\bibinfo
  {volume} {124}},\ \bibinfo {pages} {183901} (\bibinfo {year}
  {2020})}\BibitemShut {NoStop}%
\bibitem [{\citenamefont {Liu}\ \emph {et~al.}(2022)\citenamefont {Liu},
  \citenamefont {Sethi}, \citenamefont {Meng},\ and\ \citenamefont
  {Liu}}]{PhysRevB.105.085128FlatBandTPTRef3}%
  \BibitemOpen
  \bibfield  {author} {\bibinfo {author} {\bibfnamefont {H.}~\bibnamefont
  {Liu}}, \bibinfo {author} {\bibfnamefont {G.}~\bibnamefont {Sethi}}, \bibinfo
  {author} {\bibfnamefont {S.}~\bibnamefont {Meng}}, \ and\ \bibinfo {author}
  {\bibfnamefont {F.}~\bibnamefont {Liu}},\ }\href {\doibase
  10.1103/PhysRevB.105.085128} {\bibfield  {journal} {\bibinfo  {journal}
  {Phys. Rev. B}\ }\textbf {\bibinfo {volume} {105}},\ \bibinfo {pages}
  {085128} (\bibinfo {year} {2022})}\BibitemShut {NoStop}%
\bibitem [{\citenamefont {Wang}\ \emph {et~al.}(2021)\citenamefont {Wang},
  \citenamefont {Qian}, \citenamefont {Yang}, \citenamefont {Chen},
  \citenamefont {Li}, \citenamefont {Tan}, \citenamefont {Cai}, \citenamefont
  {Zhao}, \citenamefont {Gao}, \citenamefont {Feng}, \citenamefont {Kumar},
  \citenamefont {Schwier}, \citenamefont {Zhao}, \citenamefont {Weng},
  \citenamefont {Shi}, \citenamefont {Wang}, \citenamefont {Song},
  \citenamefont {Huang}, \citenamefont {Shimada}, \citenamefont {Xu},
  \citenamefont {Zhou},\ and\ \citenamefont {Liu}}]{PhysRevB.103.125131}%
  \BibitemOpen
  \bibfield  {author} {\bibinfo {author} {\bibfnamefont {Y.}~\bibnamefont
  {Wang}}, \bibinfo {author} {\bibfnamefont {Y.}~\bibnamefont {Qian}}, \bibinfo
  {author} {\bibfnamefont {M.}~\bibnamefont {Yang}}, \bibinfo {author}
  {\bibfnamefont {H.}~\bibnamefont {Chen}}, \bibinfo {author} {\bibfnamefont
  {C.}~\bibnamefont {Li}}, \bibinfo {author} {\bibfnamefont {Z.}~\bibnamefont
  {Tan}}, \bibinfo {author} {\bibfnamefont {Y.}~\bibnamefont {Cai}}, \bibinfo
  {author} {\bibfnamefont {W.}~\bibnamefont {Zhao}}, \bibinfo {author}
  {\bibfnamefont {S.}~\bibnamefont {Gao}}, \bibinfo {author} {\bibfnamefont
  {Y.}~\bibnamefont {Feng}}, \bibinfo {author} {\bibfnamefont {S.}~\bibnamefont
  {Kumar}}, \bibinfo {author} {\bibfnamefont {E.~F.}\ \bibnamefont {Schwier}},
  \bibinfo {author} {\bibfnamefont {L.}~\bibnamefont {Zhao}}, \bibinfo {author}
  {\bibfnamefont {H.}~\bibnamefont {Weng}}, \bibinfo {author} {\bibfnamefont
  {Y.}~\bibnamefont {Shi}}, \bibinfo {author} {\bibfnamefont {G.}~\bibnamefont
  {Wang}}, \bibinfo {author} {\bibfnamefont {Y.}~\bibnamefont {Song}}, \bibinfo
  {author} {\bibfnamefont {Y.}~\bibnamefont {Huang}}, \bibinfo {author}
  {\bibfnamefont {K.}~\bibnamefont {Shimada}}, \bibinfo {author} {\bibfnamefont
  {Z.}~\bibnamefont {Xu}}, \bibinfo {author} {\bibfnamefont {X.~J.}\
  \bibnamefont {Zhou}}, \ and\ \bibinfo {author} {\bibfnamefont
  {G.}~\bibnamefont {Liu}},\ }\href {\doibase 10.1103/PhysRevB.103.125131}
  {\bibfield  {journal} {\bibinfo  {journal} {Phys. Rev. B}\ }\textbf {\bibinfo
  {volume} {103}},\ \bibinfo {pages} {125131} (\bibinfo {year}
  {2021})}\BibitemShut {NoStop}%
\bibitem [{\citenamefont {Klemenz}\ \emph {et~al.}(2019)\citenamefont
  {Klemenz}, \citenamefont {Lei},\ and\ \citenamefont {Schoop}}]{2019Klemenz}%
  \BibitemOpen
  \bibfield  {author} {\bibinfo {author} {\bibfnamefont {S.}~\bibnamefont
  {Klemenz}}, \bibinfo {author} {\bibfnamefont {S.}~\bibnamefont {Lei}}, \ and\
  \bibinfo {author} {\bibfnamefont {L.~M.}\ \bibnamefont {Schoop}},\ }\href
  {\doibase 10.1146/annurev-matsci-070218-010114} {\bibfield  {journal}
  {\bibinfo  {journal} {Annu. Rev. Mater. Res.}\ }\textbf
  {\bibinfo {volume} {49}},\ \bibinfo {pages} {185} (\bibinfo {year}
  {2019})}\BibitemShut {NoStop}%
\bibitem [{\citenamefont {Wang}\ \emph {et~al.}(2012)\citenamefont {Wang},
  \citenamefont {Wang},\ and\ \citenamefont {Petrovic}}]{apl112111}%
  \BibitemOpen
  \bibfield  {author} {\bibinfo {author} {\bibfnamefont {K.}~\bibnamefont
  {Wang}}, \bibinfo {author} {\bibfnamefont {L.}~\bibnamefont {Wang}}, \ and\
  \bibinfo {author} {\bibfnamefont {C.}~\bibnamefont {Petrovic}},\ }\href
  {\doibase 10.1063/1.3695155} {\bibfield  {journal} {\bibinfo  {journal}
  {Appl. Phys. Lett.}\ }\textbf {\bibinfo {volume} {100}},\ \bibinfo
  {pages} {112111} (\bibinfo {year} {2012})}
  \BibitemShut {NoStop}%
\bibitem [{\citenamefont {Lee}\ \emph {et~al.}(2013)\citenamefont {Lee},
  \citenamefont {Farhan}, \citenamefont {Kim},\ and\ \citenamefont
  {Shim}}]{PhysRevB.87.245104}%
  \BibitemOpen
  \bibfield  {author} {\bibinfo {author} {\bibfnamefont {G.}~\bibnamefont
  {Lee}}, \bibinfo {author} {\bibfnamefont {M.~A.}\ \bibnamefont {Farhan}},
  \bibinfo {author} {\bibfnamefont {J.~S.}\ \bibnamefont {Kim}}, \ and\
  \bibinfo {author} {\bibfnamefont {J.~H.}\ \bibnamefont {Shim}},\ }\href
  {\doibase 10.1103/PhysRevB.87.245104} {\bibfield  {journal} {\bibinfo
  {journal} {Phys. Rev. B}\ }\textbf {\bibinfo {volume} {87}},\ \bibinfo
  {pages} {245104} (\bibinfo {year} {2013})}\BibitemShut {NoStop}%
\bibitem [{\citenamefont {Liu}\ \emph {et~al.}(2017)\citenamefont {Liu},
  \citenamefont {Hu}, \citenamefont {Zhang}, \citenamefont {Graf},
  \citenamefont {Cao}, \citenamefont {Radmanesh}, \citenamefont {Adams},
  \citenamefont {Zhu}, \citenamefont {Cheng}, \citenamefont {Liu},
  \citenamefont {Phelan}, \citenamefont {Wei}, \citenamefont {Jaime},
  \citenamefont {Balakirev}, \citenamefont {Tennant}, \citenamefont {DiTusa},
  \citenamefont {Chiorescu}, \citenamefont {Spinu},\ and\ \citenamefont
  {Mao}}]{Liu2017}%
  \BibitemOpen
  \bibfield  {author} {\bibinfo {author} {\bibfnamefont {J.~Y.}\ \bibnamefont
  {Liu}}, \bibinfo {author} {\bibfnamefont {J.}~\bibnamefont {Hu}}, \bibinfo
  {author} {\bibfnamefont {Q.}~\bibnamefont {Zhang}}, \bibinfo {author}
  {\bibfnamefont {D.}~\bibnamefont {Graf}}, \bibinfo {author} {\bibfnamefont
  {H.~B.}\ \bibnamefont {Cao}}, \bibinfo {author} {\bibfnamefont {S.~M.~A.}\
  \bibnamefont {Radmanesh}}, \bibinfo {author} {\bibfnamefont {D.~J.}\
  \bibnamefont {Adams}}, \bibinfo {author} {\bibfnamefont {Y.~L.}\ \bibnamefont
  {Zhu}}, \bibinfo {author} {\bibfnamefont {G.~.~F.}\ \bibnamefont {Cheng}},
  \bibinfo {author} {\bibfnamefont {X.}~\bibnamefont {Liu}}, \bibinfo {author}
  {\bibfnamefont {W.~A.}\ \bibnamefont {Phelan}}, \bibinfo {author}
  {\bibfnamefont {J.}~\bibnamefont {Wei}}, \bibinfo {author} {\bibfnamefont
  {M.}~\bibnamefont {Jaime}}, \bibinfo {author} {\bibfnamefont
  {F.}~\bibnamefont {Balakirev}}, \bibinfo {author} {\bibfnamefont {D.~A.}\
  \bibnamefont {Tennant}}, \bibinfo {author} {\bibfnamefont {J.~F.}\
  \bibnamefont {DiTusa}}, \bibinfo {author} {\bibfnamefont {I.}~\bibnamefont
  {Chiorescu}}, \bibinfo {author} {\bibfnamefont {L.}~\bibnamefont {Spinu}}, \
  and\ \bibinfo {author} {\bibfnamefont {Z.~Q.}\ \bibnamefont {Mao}},\ }\href
  {\doibase 10.1038/nmat4953} {\bibfield  {journal} {\bibinfo  {journal}
  {Nat. Mater.}\ }\textbf {\bibinfo {volume} {16}},\ \bibinfo {pages}
  {905} (\bibinfo {year} {2017})}\BibitemShut {NoStop}%
\bibitem [{\citenamefont {Liu}\ \emph {et~al.}(2021)\citenamefont {Liu},
  \citenamefont {Yu}, \citenamefont {Ning}, \citenamefont {Yi}, \citenamefont
  {Miao}, \citenamefont {Min}, \citenamefont {Zhao}, \citenamefont {Ning},
  \citenamefont {Lopez}, \citenamefont {Zhu}, \citenamefont {Pillsbury},
  \citenamefont {Zhang}, \citenamefont {Wang}, \citenamefont {Hu},
  \citenamefont {Cao}, \citenamefont {Chakoumakos}, \citenamefont {Balakirev},
  \citenamefont {Weickert}, \citenamefont {Jaime}, \citenamefont {Lai},
  \citenamefont {Yang}, \citenamefont {Sun}, \citenamefont {Alem},
  \citenamefont {Gopalan}, \citenamefont {Chang}, \citenamefont {Samarth},
  \citenamefont {Liu}, \citenamefont {McDonald},\ and\ \citenamefont
  {Mao}}]{Liu2021}%
  \BibitemOpen
  \bibfield  {author} {\bibinfo {author} {\bibfnamefont {J.~Y.}\ \bibnamefont
  {Liu}}, \bibinfo {author} {\bibfnamefont {J.}~\bibnamefont {Yu}}, \bibinfo
  {author} {\bibfnamefont {J.~L.}\ \bibnamefont {Ning}}, \bibinfo {author}
  {\bibfnamefont {H.~M.}\ \bibnamefont {Yi}}, \bibinfo {author} {\bibfnamefont
  {L.}~\bibnamefont {Miao}}, \bibinfo {author} {\bibfnamefont {L.~J.}\
  \bibnamefont {Min}}, \bibinfo {author} {\bibfnamefont {Y.~F.}\ \bibnamefont
  {Zhao}}, \bibinfo {author} {\bibfnamefont {W.}~\bibnamefont {Ning}}, \bibinfo
  {author} {\bibfnamefont {K.~A.}\ \bibnamefont {Lopez}}, \bibinfo {author}
  {\bibfnamefont {Y.~L.}\ \bibnamefont {Zhu}}, \bibinfo {author} {\bibfnamefont
  {T.}~\bibnamefont {Pillsbury}}, \bibinfo {author} {\bibfnamefont {Y.~B.}\
  \bibnamefont {Zhang}}, \bibinfo {author} {\bibfnamefont {Y.}~\bibnamefont
  {Wang}}, \bibinfo {author} {\bibfnamefont {J.}~\bibnamefont {Hu}}, \bibinfo
  {author} {\bibfnamefont {H.~B.}\ \bibnamefont {Cao}}, \bibinfo {author}
  {\bibfnamefont {B.~C.}\ \bibnamefont {Chakoumakos}}, \bibinfo {author}
  {\bibfnamefont {F.}~\bibnamefont {Balakirev}}, \bibinfo {author}
  {\bibfnamefont {F.}~\bibnamefont {Weickert}}, \bibinfo {author}
  {\bibfnamefont {M.}~\bibnamefont {Jaime}}, \bibinfo {author} {\bibfnamefont
  {Y.}~\bibnamefont {Lai}}, \bibinfo {author} {\bibfnamefont {K.}~\bibnamefont
  {Yang}}, \bibinfo {author} {\bibfnamefont {J.~W.}\ \bibnamefont {Sun}},
  \bibinfo {author} {\bibfnamefont {N.}~\bibnamefont {Alem}}, \bibinfo {author}
  {\bibfnamefont {V.}~\bibnamefont {Gopalan}}, \bibinfo {author} {\bibfnamefont
  {C.~Z.}\ \bibnamefont {Chang}}, \bibinfo {author} {\bibfnamefont
  {N.}~\bibnamefont {Samarth}}, \bibinfo {author} {\bibfnamefont {C.~X.}\
  \bibnamefont {Liu}}, \bibinfo {author} {\bibfnamefont {R.~D.}\ \bibnamefont
  {McDonald}}, \ and\ \bibinfo {author} {\bibfnamefont {Z.~Q.}\ \bibnamefont
  {Mao}},\ }\href {\doibase 10.1038/s41467-021-24369-1} {\bibfield  {journal}
  {\bibinfo  {journal} {Nat. Commun.}\ }\textbf {\bibinfo {volume}
  {12}},\ \bibinfo {pages} {4062} (\bibinfo {year} {2021})}\BibitemShut
  {NoStop}%
\bibitem [{\citenamefont {Liu}\ \emph {et~al.}(2019)\citenamefont {Liu},
  \citenamefont {Liu}, \citenamefont {Gordon}, \citenamefont {Emmanouilidou},
  \citenamefont {Xing}, \citenamefont {Graf}, \citenamefont {Chakoumakos},
  \citenamefont {Wu}, \citenamefont {Cao}, \citenamefont {Dessau},
  \citenamefont {Liu},\ and\ \citenamefont {Ni}}]{PhysRevB.100.195123}%
  \BibitemOpen
  \bibfield  {author} {\bibinfo {author} {\bibfnamefont {J.}~\bibnamefont
  {Liu}}, \bibinfo {author} {\bibfnamefont {P.}~\bibnamefont {Liu}}, \bibinfo
  {author} {\bibfnamefont {K.}~\bibnamefont {Gordon}}, \bibinfo {author}
  {\bibfnamefont {E.}~\bibnamefont {Emmanouilidou}}, \bibinfo {author}
  {\bibfnamefont {J.}~\bibnamefont {Xing}}, \bibinfo {author} {\bibfnamefont
  {D.}~\bibnamefont {Graf}}, \bibinfo {author} {\bibfnamefont {B.~C.}\
  \bibnamefont {Chakoumakos}}, \bibinfo {author} {\bibfnamefont
  {Y.}~\bibnamefont {Wu}}, \bibinfo {author} {\bibfnamefont {H.}~\bibnamefont
  {Cao}}, \bibinfo {author} {\bibfnamefont {D.}~\bibnamefont {Dessau}},
  \bibinfo {author} {\bibfnamefont {Q.}~\bibnamefont {Liu}}, \ and\ \bibinfo
  {author} {\bibfnamefont {N.}~\bibnamefont {Ni}},\ }\href {\doibase
  10.1103/PhysRevB.100.195123} {\bibfield  {journal} {\bibinfo  {journal}
  {Phys. Rev. B}\ }\textbf {\bibinfo {volume} {100}},\ \bibinfo {pages}
  {195123} (\bibinfo {year} {2019})}\BibitemShut {NoStop}%
\bibitem [{\citenamefont {Huang}\ \emph {et~al.}(2017)\citenamefont {Huang},
  \citenamefont {Kim}, \citenamefont {Shelton}, \citenamefont {Plummer},\ and\
  \citenamefont {Jin}}]{pnas.1706657114}%
  \BibitemOpen
  \bibfield  {author} {\bibinfo {author} {\bibfnamefont {S.}~\bibnamefont
  {Huang}}, \bibinfo {author} {\bibfnamefont {J.}~\bibnamefont {Kim}}, \bibinfo
  {author} {\bibfnamefont {W.~A.}\ \bibnamefont {Shelton}}, \bibinfo {author}
  {\bibfnamefont {E.~W.}\ \bibnamefont {Plummer}}, \ and\ \bibinfo {author}
  {\bibfnamefont {R.}~\bibnamefont {Jin}},\ }\href {\doibase
  10.1073/pnas.1706657114} {\bibfield  {journal} {\bibinfo  {journal}
  {Proc. Natl. Acad. Sci. USA}\ }\textbf {\bibinfo
  {volume} {114}},\ \bibinfo {pages} {6256} (\bibinfo {year} {2017})}
  \BibitemShut {NoStop}%
\bibitem [{\citenamefont {Xu}\ \emph {et~al.}(2015)\citenamefont {Xu},
  \citenamefont {Song}, \citenamefont {Nie}, \citenamefont {Weng},
  \citenamefont {Fang},\ and\ \citenamefont {Dai}}]{PhysRevB.92.205310}%
  \BibitemOpen
  \bibfield  {author} {\bibinfo {author} {\bibfnamefont {Q.}~\bibnamefont
  {Xu}}, \bibinfo {author} {\bibfnamefont {Z.}~\bibnamefont {Song}}, \bibinfo
  {author} {\bibfnamefont {S.}~\bibnamefont {Nie}}, \bibinfo {author}
  {\bibfnamefont {H.}~\bibnamefont {Weng}}, \bibinfo {author} {\bibfnamefont
  {Z.}~\bibnamefont {Fang}}, \ and\ \bibinfo {author} {\bibfnamefont
  {X.}~\bibnamefont {Dai}},\ }\href {\doibase 10.1103/PhysRevB.92.205310}
  {\bibfield  {journal} {\bibinfo  {journal} {Phys. Rev. B}\ }\textbf {\bibinfo
  {volume} {92}},\ \bibinfo {pages} {205310} (\bibinfo {year}
  {2015})}\BibitemShut {NoStop}%
\bibitem [{\citenamefont {Hulliger}\ \emph {et~al.}(1977)\citenamefont
  {Hulliger}, \citenamefont {Schmelczer},\ and\ \citenamefont
  {Schwarzenbach}}]{HULLIGER1977371}%
  \BibitemOpen
  \bibfield  {author} {\bibinfo {author} {\bibfnamefont {F.}~\bibnamefont
  {Hulliger}}, \bibinfo {author} {\bibfnamefont {R.}~\bibnamefont
  {Schmelczer}}, \ and\ \bibinfo {author} {\bibfnamefont {D.}~\bibnamefont
  {Schwarzenbach}},\ }\href {\doibase
  https://doi.org/10.1016/0022-4596(77)90134-7} {\bibfield  {journal} {\bibinfo
   {journal} {J. Solid State Chem.}\ }\textbf {\bibinfo {volume}
  {21}},\ \bibinfo {pages} {371} (\bibinfo {year} {1977})}\BibitemShut
  {NoStop}%
\bibitem [{\citenamefont {{Hoistad Strauss}}\ and\ \citenamefont
  {Delp}(2003)}]{Strauss2003CrystalSO}%
  \BibitemOpen
  \bibfield  {author} {\bibinfo {author} {\bibfnamefont {L.}~\bibnamefont
  {{Hoistad Strauss}}}\ and\ \bibinfo {author} {\bibfnamefont {C.~M.}\
  \bibnamefont {Delp}},\ }\href {\doibase
  https://doi.org/10.1016/S0925-8388(02)01308-7} {\bibfield  {journal}
  {\bibinfo  {journal} {Journal of Alloys and Compounds}\ }\textbf {\bibinfo
  {volume} {353}},\ \bibinfo {pages} {143} (\bibinfo {year}
  {2003})}\BibitemShut {NoStop}%
\bibitem [{\citenamefont {Wang}\ \emph {et~al.}(2016)\citenamefont {Wang},
  \citenamefont {Alexandradinata}, \citenamefont {Cava},\ and\ \citenamefont
  {Bernevig}}]{Wang2016Hourglass}%
  \BibitemOpen
  \bibfield  {author} {\bibinfo {author} {\bibfnamefont {Z.}~\bibnamefont
  {Wang}}, \bibinfo {author} {\bibfnamefont {A.}~\bibnamefont
  {Alexandradinata}}, \bibinfo {author} {\bibfnamefont {R.~J.}\ \bibnamefont
  {Cava}}, \ and\ \bibinfo {author} {\bibfnamefont {B.~A.}\ \bibnamefont
  {Bernevig}},\ }\href@noop {} {\bibfield  {journal} {\bibinfo  {journal}
  {Nature}\ }\textbf {\bibinfo {volume} {532}},\ \bibinfo {pages} {189}
  (\bibinfo {year} {2016})}\BibitemShut {NoStop}%
\bibitem [{\citenamefont {Fu}(2011)}]{PhysRevLett.106.106802TCI}%
  \BibitemOpen
  \bibfield  {author} {\bibinfo {author} {\bibfnamefont {L.}~\bibnamefont
  {Fu}},\ }\href {\doibase 10.1103/PhysRevLett.106.106802} {\bibfield
  {journal} {\bibinfo  {journal} {Phys. Rev. Lett.}\ }\textbf {\bibinfo
  {volume} {106}},\ \bibinfo {pages} {106802} (\bibinfo {year}
  {2011})}\BibitemShut {NoStop}%
\bibitem [{\citenamefont {Xu}\ \emph {et~al.}(2012)\citenamefont {Xu},
  \citenamefont {Liu}, \citenamefont {Alidoust}, \citenamefont {Neupane},
  \citenamefont {Qian}, \citenamefont {Belopolski}, \citenamefont {Denlinger},
  \citenamefont {Wang}, \citenamefont {Lin}, \citenamefont {Wray},
  \citenamefont {Landolt}, \citenamefont {Slomski}, \citenamefont {Dil},
  \citenamefont {Marcinkova}, \citenamefont {Morosan}, \citenamefont {Gibson},
  \citenamefont {Sankar}, \citenamefont {Chou}, \citenamefont {Cava},
  \citenamefont {Bansil},\ and\ \citenamefont {Hasan}}]{TCIExp}%
  \BibitemOpen
  \bibfield  {author} {\bibinfo {author} {\bibfnamefont {S.-Y.}\ \bibnamefont
  {Xu}}, \bibinfo {author} {\bibfnamefont {C.}~\bibnamefont {Liu}}, \bibinfo
  {author} {\bibfnamefont {N.}~\bibnamefont {Alidoust}}, \bibinfo {author}
  {\bibfnamefont {M.}~\bibnamefont {Neupane}}, \bibinfo {author} {\bibfnamefont
  {D.}~\bibnamefont {Qian}}, \bibinfo {author} {\bibfnamefont {I.}~\bibnamefont
  {Belopolski}}, \bibinfo {author} {\bibfnamefont {J.~D.}\ \bibnamefont
  {Denlinger}}, \bibinfo {author} {\bibfnamefont {Y.~J.}\ \bibnamefont {Wang}},
  \bibinfo {author} {\bibfnamefont {H.}~\bibnamefont {Lin}}, \bibinfo {author}
  {\bibfnamefont {L.~A.}\ \bibnamefont {Wray}}, \bibinfo {author}
  {\bibfnamefont {G.}~\bibnamefont {Landolt}}, \bibinfo {author} {\bibfnamefont
  {B.}~\bibnamefont {Slomski}}, \bibinfo {author} {\bibfnamefont {J.~H.}\
  \bibnamefont {Dil}}, \bibinfo {author} {\bibfnamefont {A.}~\bibnamefont
  {Marcinkova}}, \bibinfo {author} {\bibfnamefont {E.}~\bibnamefont {Morosan}},
  \bibinfo {author} {\bibfnamefont {Q.}~\bibnamefont {Gibson}}, \bibinfo
  {author} {\bibfnamefont {R.}~\bibnamefont {Sankar}}, \bibinfo {author}
  {\bibfnamefont {F.~C.}\ \bibnamefont {Chou}}, \bibinfo {author}
  {\bibfnamefont {R.~J.}\ \bibnamefont {Cava}}, \bibinfo {author}
  {\bibfnamefont {A.}~\bibnamefont {Bansil}}, \ and\ \bibinfo {author}
  {\bibfnamefont {M.~Z.}\ \bibnamefont {Hasan}},\ }\href {\doibase
  10.1038/ncomms2191} {\bibfield  {journal} {\bibinfo  {journal} {Nat.
  Commun.}\ }\textbf {\bibinfo {volume} {3}},\ \bibinfo {pages} {1192}
  (\bibinfo {year} {2012})}\BibitemShut {NoStop}%
\bibitem [{\citenamefont {Ando}\ and\ \citenamefont
  {Fu}(2015)}]{annurev-conmatphys-031214-014501}%
  \BibitemOpen
  \bibfield  {author} {\bibinfo {author} {\bibfnamefont {Y.}~\bibnamefont
  {Ando}}\ and\ \bibinfo {author} {\bibfnamefont {L.}~\bibnamefont {Fu}},\
  }\href {\doibase 10.1146/annurev-conmatphys-031214-014501} {\bibfield
  {journal} {\bibinfo  {journal} {Annu. Rev. Condens. Matter Phys.}\
  }\textbf {\bibinfo {volume} {6}},\ \bibinfo {pages} {361} (\bibinfo {year}
  {2015})}
  \BibitemShut
  {NoStop}%
\bibitem [{sup()}]{supmat}%
  \BibitemOpen
  \href@noop {} {\bibinfo  {journal} {See Supplemental Material at \weburl~for
  the calculation methods, band structures of the family, the details of the TB
  models, and the calculations of the mirror Chern number and hourglass
  invariant, which includes Refs.\cite{kresse1999, GGA-PBE1996,
  Monkhorst-BZ1976}}\ }\BibitemShut {NoStop}%
\bibitem [{\citenamefont {Kresse}\ and\ \citenamefont
  {Joubert}(1999)}]{kresse1999}%
  \BibitemOpen
  \bibfield  {author} {\bibinfo {author} {\bibfnamefont {G.}~\bibnamefont
  {Kresse}}\ and\ \bibinfo {author} {\bibfnamefont {D.}~\bibnamefont
  {Joubert}},\ }\href {\doibase 10.1103/PhysRevB.59.1758} {\bibfield  {journal}
  {\bibinfo  {journal} {Phys. Rev. B}\ }\textbf {\bibinfo {volume} {59}},\
  \bibinfo {pages} {1758} (\bibinfo {year} {1999})}\BibitemShut {NoStop}%
\bibitem [{\citenamefont {Perdew}\ \emph {et~al.}(1996)\citenamefont {Perdew},
  \citenamefont {Burke},\ and\ \citenamefont {Ernzerhof}}]{GGA-PBE1996}%
  \BibitemOpen
  \bibfield  {author} {\bibinfo {author} {\bibfnamefont {J.~P.}\ \bibnamefont
  {Perdew}}, \bibinfo {author} {\bibfnamefont {K.}~\bibnamefont {Burke}}, \
  and\ \bibinfo {author} {\bibfnamefont {M.}~\bibnamefont {Ernzerhof}},\ }\href
  {\doibase 10.1103/PhysRevLett.77.3865} {\bibfield  {journal} {\bibinfo
  {journal} {Phys. Rev. Lett.}\ }\textbf {\bibinfo {volume} {77}},\ \bibinfo
  {pages} {3865} (\bibinfo {year} {1996})}\BibitemShut {NoStop}%
\bibitem [{\citenamefont {Monkhorst}\ and\ \citenamefont
  {Pack}(1976)}]{Monkhorst-BZ1976}%
  \BibitemOpen
  \bibfield  {author} {\bibinfo {author} {\bibfnamefont {H.~J.}\ \bibnamefont
  {Monkhorst}}\ and\ \bibinfo {author} {\bibfnamefont {J.~D.}\ \bibnamefont
  {Pack}},\ }\href {\doibase 10.1103/PhysRevB.13.5188} {\bibfield  {journal}
  {\bibinfo  {journal} {Phys. Rev. B}\ }\textbf {\bibinfo {volume} {13}},\
  \bibinfo {pages} {5188} (\bibinfo {year} {1976})}\BibitemShut {NoStop}%
\bibitem [{\citenamefont {Gao}\ \emph {et~al.}(2021{\natexlab{b}})\citenamefont
  {Gao}, \citenamefont {Wu}, \citenamefont {Persson},\ and\ \citenamefont
  {Wang}}]{gao2021}%
  \BibitemOpen
\bibfield  {journal} {  }\bibfield  {author} {\bibinfo {author} {\bibfnamefont
  {J.}~\bibnamefont {Gao}}, \bibinfo {author} {\bibfnamefont {Q.}~\bibnamefont
  {Wu}}, \bibinfo {author} {\bibfnamefont {C.}~\bibnamefont {Persson}}, \ and\
  \bibinfo {author} {\bibfnamefont {Z.}~\bibnamefont {Wang}},\ }\href {\doibase
  10.1016/j.cpc.2020.107760} {\bibfield  {journal} {\bibinfo  {journal}
  {Comput. Phys. Commun.}\ }\textbf {\bibinfo {volume} {261}},\
  \bibinfo {pages} {107760} (\bibinfo {year} {2021}{\natexlab{b}})}\BibitemShut
  {NoStop}%
\bibitem [{\citenamefont {Fu}\ and\ \citenamefont
  {Kane}(2007)}]{PhysRevB.76.045302FK}%
  \BibitemOpen
  \bibfield  {author} {\bibinfo {author} {\bibfnamefont {L.}~\bibnamefont
  {Fu}}\ and\ \bibinfo {author} {\bibfnamefont {C.~L.}\ \bibnamefont {Kane}},\
  }\href {\doibase 10.1103/PhysRevB.76.045302} {\bibfield  {journal} {\bibinfo
  {journal} {Phys. Rev. B}\ }\textbf {\bibinfo {volume} {76}},\ \bibinfo
  {pages} {045302} (\bibinfo {year} {2007})}\BibitemShut {NoStop}%
\bibitem [{\citenamefont {Po}\ \emph {et~al.}(2017)\citenamefont {Po},
  \citenamefont {Vishwanath},\ and\ \citenamefont
  {Watanabe}}]{SymmIndicator2017Po}%
  \BibitemOpen
  \bibfield  {author} {\bibinfo {author} {\bibfnamefont {H.~C.}\ \bibnamefont
  {Po}}, \bibinfo {author} {\bibfnamefont {A.}~\bibnamefont {Vishwanath}}, \
  and\ \bibinfo {author} {\bibfnamefont {H.}~\bibnamefont {Watanabe}},\ }\href
  {\doibase 10.1038/s41467-017-00133-2} {\bibfield  {journal} {\bibinfo
  {journal} {Nat. Commun.}\ }\textbf {\bibinfo {volume} {8}},\
  \bibinfo {pages} {50} (\bibinfo {year} {2017})}\BibitemShut {NoStop}%
\bibitem [{\citenamefont {Song}\ \emph {et~al.}(2018)\citenamefont {Song},
  \citenamefont {Zhang}, \citenamefont {Fang},\ and\ \citenamefont
  {Fang}}]{songzd2018natcomm}%
  \BibitemOpen
  \bibfield  {author} {\bibinfo {author} {\bibfnamefont {Z.}~\bibnamefont
  {Song}}, \bibinfo {author} {\bibfnamefont {T.}~\bibnamefont {Zhang}},
  \bibinfo {author} {\bibfnamefont {Z.}~\bibnamefont {Fang}}, \ and\ \bibinfo
  {author} {\bibfnamefont {C.}~\bibnamefont {Fang}},\ }\href {\doibase
  10.1038/s41467-018-06010-w} {\bibfield  {journal} {\bibinfo  {journal}
  {Nat. Commun.}\ }\textbf {\bibinfo {volume} {9}},\ \bibinfo {pages}
  {3530} (\bibinfo {year} {2018})}\BibitemShut {NoStop}%
\bibitem [{\citenamefont {Tang}\ \emph {et~al.}(2019)\citenamefont {Tang},
  \citenamefont {Po}, \citenamefont {Vishwanath},\ and\ \citenamefont
  {Wan}}]{SymmIndicatorData2019Tang}%
  \BibitemOpen
  \bibfield  {author} {\bibinfo {author} {\bibfnamefont {F.}~\bibnamefont
  {Tang}}, \bibinfo {author} {\bibfnamefont {H.~C.}\ \bibnamefont {Po}},
  \bibinfo {author} {\bibfnamefont {A.}~\bibnamefont {Vishwanath}}, \ and\
  \bibinfo {author} {\bibfnamefont {X.}~\bibnamefont {Wan}},\ }\href {\doibase
  10.1038/s41586-019-0937-5} {\bibfield  {journal} {\bibinfo  {journal}
  {Nature}\ }\textbf {\bibinfo {volume} {566}},\ \bibinfo {pages} {486}
  (\bibinfo {year} {2019})}\BibitemShut {NoStop}%
\bibitem [{\citenamefont {Zhang}\ \emph {et~al.}(2019)\citenamefont {Zhang},
  \citenamefont {Jiang}, \citenamefont {Song}, \citenamefont {Huang},
  \citenamefont {He}, \citenamefont {Fang}, \citenamefont {Weng},\ and\
  \citenamefont {Fang}}]{SymmIndi2019ZhangTT}%
  \BibitemOpen
  \bibfield  {author} {\bibinfo {author} {\bibfnamefont {T.}~\bibnamefont
  {Zhang}}, \bibinfo {author} {\bibfnamefont {Y.}~\bibnamefont {Jiang}},
  \bibinfo {author} {\bibfnamefont {Z.}~\bibnamefont {Song}}, \bibinfo {author}
  {\bibfnamefont {H.}~\bibnamefont {Huang}}, \bibinfo {author} {\bibfnamefont
  {Y.}~\bibnamefont {He}}, \bibinfo {author} {\bibfnamefont {Z.}~\bibnamefont
  {Fang}}, \bibinfo {author} {\bibfnamefont {H.}~\bibnamefont {Weng}}, \ and\
  \bibinfo {author} {\bibfnamefont {C.}~\bibnamefont {Fang}},\ }\href {\doibase
  10.1038/s41586-019-0944-6} {\bibfield  {journal} {\bibinfo  {journal}
  {Nature}\ }\textbf {\bibinfo {volume} {566}},\ \bibinfo {pages} {475}
  (\bibinfo {year} {2019})}\BibitemShut {NoStop}%
\bibitem [{\citenamefont {Vergniory}\ \emph {et~al.}(2019)\citenamefont
  {Vergniory}, \citenamefont {Elcoro}, \citenamefont {Felser}, \citenamefont
  {Regnault}, \citenamefont {Bernevig},\ and\ \citenamefont
  {Wang}}]{Vergniory2019}%
  \BibitemOpen
  \bibfield  {author} {\bibinfo {author} {\bibfnamefont {M.~G.}\ \bibnamefont
  {Vergniory}}, \bibinfo {author} {\bibfnamefont {L.}~\bibnamefont {Elcoro}},
  \bibinfo {author} {\bibfnamefont {C.}~\bibnamefont {Felser}}, \bibinfo
  {author} {\bibfnamefont {N.}~\bibnamefont {Regnault}}, \bibinfo {author}
  {\bibfnamefont {B.~A.}\ \bibnamefont {Bernevig}}, \ and\ \bibinfo {author}
  {\bibfnamefont {Z.}~\bibnamefont {Wang}},\ }\href {\doibase
  10.1038/s41586-019-0954-4} {\bibfield  {journal} {\bibinfo  {journal}
  {Nature}\ }\textbf {\bibinfo {volume} {566}},\ \bibinfo {pages} {480}
  (\bibinfo {year} {2019})}\BibitemShut {NoStop}%
\bibitem [{\citenamefont {Alexandradinata}\ \emph {et~al.}(2016)\citenamefont
  {Alexandradinata}, \citenamefont {Wang},\ and\ \citenamefont
  {Bernevig}}]{PhysRevX.6.021008}%
  \BibitemOpen
  \bibfield  {author} {\bibinfo {author} {\bibfnamefont {A.}~\bibnamefont
  {Alexandradinata}}, \bibinfo {author} {\bibfnamefont {Z.}~\bibnamefont
  {Wang}}, \ and\ \bibinfo {author} {\bibfnamefont {B.~A.}\ \bibnamefont
  {Bernevig}},\ }\href {\doibase 10.1103/PhysRevX.6.021008} {\bibfield
  {journal} {\bibinfo  {journal} {Phys. Rev. X}\ }\textbf {\bibinfo {volume}
  {6}},\ \bibinfo {pages} {021008} (\bibinfo {year} {2016})}\BibitemShut
  {NoStop}%
\bibitem [{\citenamefont {Ezawa}(2016)}]{PhysRevB.94.155148}%
  \BibitemOpen
  \bibfield  {author} {\bibinfo {author} {\bibfnamefont {M.}~\bibnamefont
  {Ezawa}},\ }\href {\doibase 10.1103/PhysRevB.94.155148} {\bibfield  {journal}
  {\bibinfo  {journal} {Phys. Rev. B}\ }\textbf {\bibinfo {volume} {94}},\
  \bibinfo {pages} {155148} (\bibinfo {year} {2016})}\BibitemShut {NoStop}%
\bibitem [{\citenamefont {Zhang}\ \emph {et~al.}(2020)\citenamefont {Zhang},
  \citenamefont {Cui}, \citenamefont {Wang}, \citenamefont {Weng},\ and\
  \citenamefont {Fang}}]{PhysRevB.101.115145}%
  \BibitemOpen
  \bibfield  {author} {\bibinfo {author} {\bibfnamefont {T.}~\bibnamefont
  {Zhang}}, \bibinfo {author} {\bibfnamefont {Z.}~\bibnamefont {Cui}}, \bibinfo
  {author} {\bibfnamefont {Z.}~\bibnamefont {Wang}}, \bibinfo {author}
  {\bibfnamefont {H.}~\bibnamefont {Weng}}, \ and\ \bibinfo {author}
  {\bibfnamefont {Z.}~\bibnamefont {Fang}},\ }\href {\doibase
  10.1103/PhysRevB.101.115145} {\bibfield  {journal} {\bibinfo  {journal}
  {Phys. Rev. B}\ }\textbf {\bibinfo {volume} {101}},\ \bibinfo {pages}
  {115145} (\bibinfo {year} {2020})}\BibitemShut {NoStop}%
\bibitem [{\citenamefont {Slater}\ and\ \citenamefont
  {Koster}(1954)}]{PhysRev.94.1498LCAO}%
  \BibitemOpen
  \bibfield  {author} {\bibinfo {author} {\bibfnamefont {J.~C.}\ \bibnamefont
  {Slater}}\ and\ \bibinfo {author} {\bibfnamefont {G.~F.}\ \bibnamefont
  {Koster}},\ }\href {\doibase 10.1103/PhysRev.94.1498} {\bibfield  {journal}
  {\bibinfo  {journal} {Phys. Rev.}\ }\textbf {\bibinfo {volume} {94}},\
  \bibinfo {pages} {1498} (\bibinfo {year} {1954})}\BibitemShut {NoStop}%
\bibitem [{\citenamefont {Berry}(1984)}]{rspa.1984.0023}%
  \BibitemOpen
  \bibfield  {author} {\bibinfo {author} {\bibfnamefont {M.~V.}\ \bibnamefont
  {Berry}},\ }\href {\doibase 10.1098/rspa.1984.0023} {\bibfield  {journal}
  {\bibinfo  {journal} {Proc. R. Soc. London. Ser. A}\ }\textbf {\bibinfo {volume} {392}},\
  \bibinfo {pages} {45} (\bibinfo {year} {1984})}
  \BibitemShut {NoStop}%
\bibitem [{\citenamefont {Zak}(1989)}]{PhysRevLett.62.2747}%
  \BibitemOpen
  \bibfield  {author} {\bibinfo {author} {\bibfnamefont {J.}~\bibnamefont
  {Zak}},\ }\href {\doibase 10.1103/PhysRevLett.62.2747} {\bibfield  {journal}
  {\bibinfo  {journal} {Phys. Rev. Lett.}\ }\textbf {\bibinfo {volume} {62}},\
  \bibinfo {pages} {2747} (\bibinfo {year} {1989})}\BibitemShut {NoStop}%
\bibitem [{\citenamefont {Yu}\ \emph {et~al.}(2011)\citenamefont {Yu},
  \citenamefont {Qi}, \citenamefont {Bernevig}, \citenamefont {Fang},\ and\
  \citenamefont {Dai}}]{PhysRevB.84.075119}%
  \BibitemOpen
  \bibfield  {author} {\bibinfo {author} {\bibfnamefont {R.}~\bibnamefont
  {Yu}}, \bibinfo {author} {\bibfnamefont {X.~L.}\ \bibnamefont {Qi}}, \bibinfo
  {author} {\bibfnamefont {A.}~\bibnamefont {Bernevig}}, \bibinfo {author}
  {\bibfnamefont {Z.}~\bibnamefont {Fang}}, \ and\ \bibinfo {author}
  {\bibfnamefont {X.}~\bibnamefont {Dai}},\ }\href {\doibase
  10.1103/PhysRevB.84.075119} {\bibfield  {journal} {\bibinfo  {journal} {Phys.
  Rev. B}\ }\textbf {\bibinfo {volume} {84}},\ \bibinfo {pages} {075119}
  (\bibinfo {year} {2011})}\BibitemShut {NoStop}%
\bibitem [{\citenamefont {Alexandradinata}\ \emph {et~al.}(2014)\citenamefont
  {Alexandradinata}, \citenamefont {Dai},\ and\ \citenamefont
  {Bernevig}}]{PhysRevB.89.155114}%
  \BibitemOpen
  \bibfield  {author} {\bibinfo {author} {\bibfnamefont {A.}~\bibnamefont
  {Alexandradinata}}, \bibinfo {author} {\bibfnamefont {X.}~\bibnamefont
  {Dai}}, \ and\ \bibinfo {author} {\bibfnamefont {B.~A.}\ \bibnamefont
  {Bernevig}},\ }\href {\doibase 10.1103/PhysRevB.89.155114} {\bibfield
  {journal} {\bibinfo  {journal} {Phys. Rev. B}\ }\textbf {\bibinfo {volume}
  {89}},\ \bibinfo {pages} {155114} (\bibinfo {year} {2014})}\BibitemShut
  {NoStop}%
\bibitem [{\citenamefont {Alexandradinata}\ and\ \citenamefont
  {Bernevig}(2016)}]{PhysRevB.93.205104}%
  \BibitemOpen
  \bibfield  {author} {\bibinfo {author} {\bibfnamefont {A.}~\bibnamefont
  {Alexandradinata}}\ and\ \bibinfo {author} {\bibfnamefont {B.~A.}\
  \bibnamefont {Bernevig}},\ }\href {\doibase 10.1103/PhysRevB.93.205104}
  {\bibfield  {journal} {\bibinfo  {journal} {Phys. Rev. B}\ }\textbf {\bibinfo
  {volume} {93}},\ \bibinfo {pages} {205104} (\bibinfo {year}
  {2016})}\BibitemShut {NoStop}%
\bibitem [{\citenamefont {Fidkowski}\ \emph {et~al.}(2011)\citenamefont
  {Fidkowski}, \citenamefont {Jackson},\ and\ \citenamefont
  {Klich}}]{PhysRevLett.107.036601}%
  \BibitemOpen
  \bibfield  {author} {\bibinfo {author} {\bibfnamefont {L.}~\bibnamefont
  {Fidkowski}}, \bibinfo {author} {\bibfnamefont {T.~S.}\ \bibnamefont
  {Jackson}}, \ and\ \bibinfo {author} {\bibfnamefont {I.}~\bibnamefont
  {Klich}},\ }\href {\doibase 10.1103/PhysRevLett.107.036601} {\bibfield
  {journal} {\bibinfo  {journal} {Phys. Rev. Lett.}\ }\textbf {\bibinfo
  {volume} {107}},\ \bibinfo {pages} {036601} (\bibinfo {year}
  {2011})}\BibitemShut {NoStop}%
\bibitem [{\citenamefont {Wu}\ \emph {et~al.}(2018)\citenamefont {Wu},
  \citenamefont {Zhang}, \citenamefont {Song}, \citenamefont {Troyer},\ and\
  \citenamefont {Soluyanov}}]{WU2017}%
  \BibitemOpen
  \bibfield  {author} {\bibinfo {author} {\bibfnamefont {Q.}~\bibnamefont
  {Wu}}, \bibinfo {author} {\bibfnamefont {S.}~\bibnamefont {Zhang}}, \bibinfo
  {author} {\bibfnamefont {H.-F.}\ \bibnamefont {Song}}, \bibinfo {author}
  {\bibfnamefont {M.}~\bibnamefont {Troyer}}, \ and\ \bibinfo {author}
  {\bibfnamefont {A.~A.}\ \bibnamefont {Soluyanov}},\ }\href {\doibase
  https://doi.org/10.1016/j.cpc.2017.09.033} {\bibfield  {journal} {\bibinfo
  {journal} {Comput. Phys. Commun.}\ }\textbf {\bibinfo {volume}
  {224}},\ \bibinfo {pages} {405 } (\bibinfo {year} {2018})}\BibitemShut
  {NoStop}%
\bibitem [{\citenamefont {Wu}\ \emph {et~al.}(2015)\citenamefont {Wu},
  \citenamefont {Qin}, \citenamefont {Liang}, \citenamefont {Le}, \citenamefont
  {Fan},\ and\ \citenamefont {Hu}}]{PhysRevB.91.081111}%
  \BibitemOpen
  \bibfield  {author} {\bibinfo {author} {\bibfnamefont {X.}~\bibnamefont
  {Wu}}, \bibinfo {author} {\bibfnamefont {S.}~\bibnamefont {Qin}}, \bibinfo
  {author} {\bibfnamefont {Y.}~\bibnamefont {Liang}}, \bibinfo {author}
  {\bibfnamefont {C.}~\bibnamefont {Le}}, \bibinfo {author} {\bibfnamefont
  {H.}~\bibnamefont {Fan}}, \ and\ \bibinfo {author} {\bibfnamefont
  {J.}~\bibnamefont {Hu}},\ }\href {\doibase 10.1103/PhysRevB.91.081111}
  {\bibfield  {journal} {\bibinfo  {journal} {Phys. Rev. B}\ }\textbf {\bibinfo
  {volume} {91}},\ \bibinfo {pages} {081111} (\bibinfo {year}
  {2015})}\BibitemShut {NoStop}%
\bibitem [{\citenamefont {Qian}\ \emph
  {et~al.}(2020{\natexlab{b}})\citenamefont {Qian}, \citenamefont {Tan},
  \citenamefont {Zhang}, \citenamefont {Gao}, \citenamefont {Wang},
  \citenamefont {Fang}, \citenamefont {Fang},\ and\ \citenamefont
  {Weng}}]{Qian2020}%
  \BibitemOpen
  \bibfield  {author} {\bibinfo {author} {\bibfnamefont {Y.}~\bibnamefont
  {Qian}}, \bibinfo {author} {\bibfnamefont {Z.}~\bibnamefont {Tan}}, \bibinfo
  {author} {\bibfnamefont {T.}~\bibnamefont {Zhang}}, \bibinfo {author}
  {\bibfnamefont {J.}~\bibnamefont {Gao}}, \bibinfo {author} {\bibfnamefont
  {Z.}~\bibnamefont {Wang}}, \bibinfo {author} {\bibfnamefont {Z.}~\bibnamefont
  {Fang}}, \bibinfo {author} {\bibfnamefont {C.}~\bibnamefont {Fang}}, \ and\
  \bibinfo {author} {\bibfnamefont {H.}~\bibnamefont {Weng}},\ }\href {\doibase
  10.1007/s11433-019-1515-4} {\bibfield  {journal} {\bibinfo  {journal}
  {Sci. China Phys. Mech. Astron.}\ }\textbf {\bibinfo
  {volume} {63}} (\bibinfo {year} {2020})}
  \BibitemShut {NoStop}%
\bibitem [{\citenamefont {Schlechte}\ \emph {et~al.}(2009)\citenamefont
  {Schlechte}, \citenamefont {Niewa}, \citenamefont {Prots}, \citenamefont
  {Schnelle}, \citenamefont {Schmidt},\ and\ \citenamefont
  {Kniep}}]{CeAsSe_Synthesis}%
  \BibitemOpen
  \bibfield  {author} {\bibinfo {author} {\bibfnamefont {A.}~\bibnamefont
  {Schlechte}}, \bibinfo {author} {\bibfnamefont {R.}~\bibnamefont {Niewa}},
  \bibinfo {author} {\bibfnamefont {Y.}~\bibnamefont {Prots}}, \bibinfo
  {author} {\bibfnamefont {W.}~\bibnamefont {Schnelle}}, \bibinfo {author}
  {\bibfnamefont {M.}~\bibnamefont {Schmidt}}, \ and\ \bibinfo {author}
  {\bibfnamefont {R.}~\bibnamefont {Kniep}},\ }\href {\doibase
  10.1021/ic802246g} {\bibfield  {journal} {\bibinfo  {journal} {Inorg.
  Chem.} \textbf{\bibinfo {volume} {48}},\ \ \bibinfo {pages} {2277}} (\bibinfo {year} {2009})}\BibitemShut
  {NoStop}%
\bibitem [{\citenamefont {Ohno}\ \emph {et~al.}(2021)\citenamefont {Ohno},
  \citenamefont {Uchida}, \citenamefont {Nakazawa}, \citenamefont {Sato},
  \citenamefont {Kriener}, \citenamefont {Miyake}, \citenamefont {Tokunaga},
  \citenamefont {Taguchi},\ and\ \citenamefont {Kawasaki}}]{Ohno_2021}%
  \BibitemOpen
  \bibfield  {author} {\bibinfo {author} {\bibfnamefont {M.}~\bibnamefont
  {Ohno}}, \bibinfo {author} {\bibfnamefont {M.}~\bibnamefont {Uchida}},
  \bibinfo {author} {\bibfnamefont {Y.}~\bibnamefont {Nakazawa}}, \bibinfo
  {author} {\bibfnamefont {S.}~\bibnamefont {Sato}}, \bibinfo {author}
  {\bibfnamefont {M.}~\bibnamefont {Kriener}}, \bibinfo {author} {\bibfnamefont
  {A.}~\bibnamefont {Miyake}}, \bibinfo {author} {\bibfnamefont
  {M.}~\bibnamefont {Tokunaga}}, \bibinfo {author} {\bibfnamefont
  {Y.}~\bibnamefont {Taguchi}}, \ and\ \bibinfo {author} {\bibfnamefont
  {M.}~\bibnamefont {Kawasaki}},\ }\href {\doibase 10.1063/5.0043453}
  {\bibfield  {journal} {\bibinfo  {journal} {APL Mater.}\ }\textbf
  {\bibinfo {volume} {9}},\ \bibinfo {pages} {051107} (\bibinfo {year}
  {2021})}\BibitemShut {NoStop}%
\bibitem [{\citenamefont {Yang}\ \emph {et~al.}(2011)\citenamefont {Yang},
  \citenamefont {Xu}, \citenamefont {Sheng}, \citenamefont {Wang},
  \citenamefont {Xing},\ and\ \citenamefont {Sheng}}]{PhysRevLett.107.066602}%
  \BibitemOpen
  \bibfield  {author} {\bibinfo {author} {\bibfnamefont {Y.}~\bibnamefont
  {Yang}}, \bibinfo {author} {\bibfnamefont {Z.}~\bibnamefont {Xu}}, \bibinfo
  {author} {\bibfnamefont {L.}~\bibnamefont {Sheng}}, \bibinfo {author}
  {\bibfnamefont {B.}~\bibnamefont {Wang}}, \bibinfo {author} {\bibfnamefont
  {D.~Y.}\ \bibnamefont {Xing}}, \ and\ \bibinfo {author} {\bibfnamefont
  {D.~N.}\ \bibnamefont {Sheng}},\ }\href {\doibase
  10.1103/PhysRevLett.107.066602} {\bibfield  {journal} {\bibinfo  {journal}
  {Phys. Rev. Lett.}\ }\textbf {\bibinfo {volume} {107}},\ \bibinfo {pages}
  {066602} (\bibinfo {year} {2011})}\BibitemShut {NoStop}%
\bibitem [{\citenamefont {Zhang}\ \emph {et~al.}(2017)\citenamefont {Zhang},
  \citenamefont {Sun}, \citenamefont {Yang}, \citenamefont
  {\ifmmode~\check{Z}\else \v{Z}\fi{}elezn\'y}, \citenamefont {Parkin},
  \citenamefont {Felser},\ and\ \citenamefont {Yan}}]{zhang2017}%
  \BibitemOpen
  \bibfield  {author} {\bibinfo {author} {\bibfnamefont {Y.}~\bibnamefont
  {Zhang}}, \bibinfo {author} {\bibfnamefont {Y.}~\bibnamefont {Sun}}, \bibinfo
  {author} {\bibfnamefont {H.}~\bibnamefont {Yang}}, \bibinfo {author}
  {\bibfnamefont {J.}~\bibnamefont {\ifmmode~\check{Z}\else
  \v{Z}\fi{}elezn\'y}}, \bibinfo {author} {\bibfnamefont {S.~P.~P.}\
  \bibnamefont {Parkin}}, \bibinfo {author} {\bibfnamefont {C.}~\bibnamefont
  {Felser}}, \ and\ \bibinfo {author} {\bibfnamefont {B.}~\bibnamefont {Yan}},\
  }\href {\doibase 10.1103/PhysRevB.95.075128} {\bibfield  {journal} {\bibinfo
  {journal} {Phys. Rev. B}\ }\textbf {\bibinfo {volume} {95}},\ \bibinfo
  {pages} {075128} (\bibinfo {year} {2017})}\BibitemShut {NoStop}%
\bibitem [{\citenamefont {Xiao}\ \emph {et~al.}(2010)\citenamefont {Xiao},
  \citenamefont {Chang},\ and\ \citenamefont {Niu}}]{RevModPhys.82.1959}%
  \BibitemOpen
  \bibfield  {author} {\bibinfo {author} {\bibfnamefont {D.}~\bibnamefont
  {Xiao}}, \bibinfo {author} {\bibfnamefont {M.-C.}\ \bibnamefont {Chang}}, \
  and\ \bibinfo {author} {\bibfnamefont {Q.}~\bibnamefont {Niu}},\ }\href
  {\doibase 10.1103/RevModPhys.82.1959} {\bibfield  {journal} {\bibinfo
  {journal} {Rev. Mod. Phys.}\ }\textbf {\bibinfo {volume} {82}},\ \bibinfo
  {pages} {1959} (\bibinfo {year} {2010})}\BibitemShut {NoStop}%
\bibitem [{\citenamefont {King-Smith}\ and\ \citenamefont
  {Vanderbilt}(1993)}]{PhysRevB.47.1651}%
  \BibitemOpen
  \bibfield  {author} {\bibinfo {author} {\bibfnamefont {R.~D.}\ \bibnamefont
  {King-Smith}}\ and\ \bibinfo {author} {\bibfnamefont {D.}~\bibnamefont
  {Vanderbilt}},\ }\href {\doibase 10.1103/PhysRevB.47.1651} {\bibfield
  {journal} {\bibinfo  {journal} {Phys. Rev. B}\ }\textbf {\bibinfo {volume}
  {47}},\ \bibinfo {pages} {1651} (\bibinfo {year} {1993})}\BibitemShut
  {NoStop}%
\bibitem [{\citenamefont {Vanderbilt}\ and\ \citenamefont
  {King-Smith}(1993)}]{PhysRevB.48.4442}%
  \BibitemOpen
  \bibfield  {author} {\bibinfo {author} {\bibfnamefont {D.}~\bibnamefont
  {Vanderbilt}}\ and\ \bibinfo {author} {\bibfnamefont {R.~D.}\ \bibnamefont
  {King-Smith}},\ }\href {\doibase 10.1103/PhysRevB.48.4442} {\bibfield
  {journal} {\bibinfo  {journal} {Phys. Rev. B}\ }\textbf {\bibinfo {volume}
  {48}},\ \bibinfo {pages} {4442} (\bibinfo {year} {1993})}\BibitemShut
  {NoStop}%
\bibitem [{\citenamefont {Vanderbilt}(2018)}]{berry_phase_David}%
  \BibitemOpen
  \bibfield  {author} {\bibinfo {author} {\bibfnamefont {D.}~\bibnamefont
  {Vanderbilt}},\ }\href {\doibase DOI: 10.1017/9781316662205} {\emph {\bibinfo
  {title} {Berry Phases in Electronic Structure Theory: Electric Polarization,
  Orbital Magnetization and Topological Insulators}}}\ (\bibinfo  {publisher}
  {Cambridge University Press},\ \bibinfo {address} {Cambridge},\ \bibinfo
  {year} {2018})\BibitemShut {NoStop}%
\bibitem [{\citenamefont {Katayama}\ \emph {et~al.}(2013)\citenamefont
  {Katayama}, \citenamefont {Kudo}, \citenamefont {Onari}, \citenamefont
  {Mizukami}, \citenamefont {Sugawara}, \citenamefont {Sugiyama}, \citenamefont
  {Kitahama}, \citenamefont {Iba}, \citenamefont {Fujimura}, \citenamefont
  {Nishimoto}, \citenamefont {Nohara},\ and\ \citenamefont
  {Sawa}}]{supr-con_112_jpn}%
  \BibitemOpen
  \bibfield  {author} {\bibinfo {author} {\bibfnamefont {N.}~\bibnamefont
  {Katayama}}, \bibinfo {author} {\bibfnamefont {K.}~\bibnamefont {Kudo}},
  \bibinfo {author} {\bibfnamefont {S.}~\bibnamefont {Onari}}, \bibinfo
  {author} {\bibfnamefont {T.}~\bibnamefont {Mizukami}}, \bibinfo {author}
  {\bibfnamefont {K.}~\bibnamefont {Sugawara}}, \bibinfo {author}
  {\bibfnamefont {Y.}~\bibnamefont {Sugiyama}}, \bibinfo {author}
  {\bibfnamefont {Y.}~\bibnamefont {Kitahama}}, \bibinfo {author}
  {\bibfnamefont {K.}~\bibnamefont {Iba}}, \bibinfo {author} {\bibfnamefont
  {K.}~\bibnamefont {Fujimura}}, \bibinfo {author} {\bibfnamefont
  {N.}~\bibnamefont {Nishimoto}}, \bibinfo {author} {\bibfnamefont
  {M.}~\bibnamefont {Nohara}}, \ and\ \bibinfo {author} {\bibfnamefont
  {H.}~\bibnamefont {Sawa}},\ }\href {\doibase 10.7566/JPSJ.82.123702}
  {\bibfield  {journal} {\bibinfo  {journal} {J. Phys. Soc. Jpn.}\ }\textbf {\bibinfo {volume} {82}},\ \bibinfo {pages} {123702}
  (\bibinfo {year} {2013})}
  \BibitemShut {NoStop}%
\bibitem [{\citenamefont {Takahashi}\ \emph {et~al.}(2021)\citenamefont
  {Takahashi}, \citenamefont {Kitagawa}, \citenamefont {Ishida}, \citenamefont
  {Kawaguchi}, \citenamefont {Ikeda}, \citenamefont {Yonezawa},\ and\
  \citenamefont {Maeno}}]{supr-con_ax2_jpn}%
  \BibitemOpen
  \bibfield  {author} {\bibinfo {author} {\bibfnamefont {H.}~\bibnamefont
  {Takahashi}}, \bibinfo {author} {\bibfnamefont {S.}~\bibnamefont {Kitagawa}},
  \bibinfo {author} {\bibfnamefont {K.}~\bibnamefont {Ishida}}, \bibinfo
  {author} {\bibfnamefont {M.}~\bibnamefont {Kawaguchi}}, \bibinfo {author}
  {\bibfnamefont {A.}~\bibnamefont {Ikeda}}, \bibinfo {author} {\bibfnamefont
  {S.}~\bibnamefont {Yonezawa}}, \ and\ \bibinfo {author} {\bibfnamefont
  {Y.}~\bibnamefont {Maeno}},\ }\href {\doibase 10.7566/JPSJ.90.073702}
  {\bibfield  {journal} {\bibinfo  {journal} {J. Phys. Soc. Jpn.}\ }\textbf {\bibinfo {volume} {90}},\ \bibinfo {pages} {073702}
  (\bibinfo {year} {2021})}
  \BibitemShut {NoStop}%
\bibitem [{\citenamefont {Funada}\ \emph {et~al.}(2019)\citenamefont {Funada},
  \citenamefont {Yamakage}, \citenamefont {Yamashina},\ and\ \citenamefont
  {Kageyama}}]{supr-con_ax2_2_jpn}%
  \BibitemOpen
  \bibfield  {author} {\bibinfo {author} {\bibfnamefont {K.}~\bibnamefont
  {Funada}}, \bibinfo {author} {\bibfnamefont {A.}~\bibnamefont {Yamakage}},
  \bibinfo {author} {\bibfnamefont {N.}~\bibnamefont {Yamashina}}, \ and\
  \bibinfo {author} {\bibfnamefont {H.}~\bibnamefont {Kageyama}},\ }\href
  {\doibase 10.7566/JPSJ.88.044711} {\bibfield  {journal} {\bibinfo  {journal}
  {J. Phys. Soc. Jpn.}\ }\textbf {\bibinfo {volume}
  {88}},\ \bibinfo {pages} {044711} (\bibinfo {year} {2019})}
  \BibitemShut {NoStop}%
\bibitem [{\citenamefont {Wu}\ \emph {et~al.}(2020)\citenamefont {Wu},
  \citenamefont {Benalcazar}, \citenamefont {Li}, \citenamefont {Thomale},
  \citenamefont {Liu},\ and\ \citenamefont {Hu}}]{PhysRevX.10.041014}%
  \BibitemOpen
  \bibfield  {author} {\bibinfo {author} {\bibfnamefont {X.}~\bibnamefont
  {Wu}}, \bibinfo {author} {\bibfnamefont {W.~A.}\ \bibnamefont {Benalcazar}},
  \bibinfo {author} {\bibfnamefont {Y.}~\bibnamefont {Li}}, \bibinfo {author}
  {\bibfnamefont {R.}~\bibnamefont {Thomale}}, \bibinfo {author} {\bibfnamefont
  {C.-X.}\ \bibnamefont {Liu}}, \ and\ \bibinfo {author} {\bibfnamefont
  {J.}~\bibnamefont {Hu}},\ }\href {\doibase 10.1103/PhysRevX.10.041014}
  {\bibfield  {journal} {\bibinfo  {journal} {Phys. Rev. X}\ }\textbf {\bibinfo
  {volume} {10}},\ \bibinfo {pages} {041014} (\bibinfo {year}
  {2020})}\BibitemShut {NoStop}%
\bibitem [{\citenamefont {Wang}\ \emph {et~al.}(2015)\citenamefont {Wang},
  \citenamefont {Zhang}, \citenamefont {Xu}, \citenamefont {Zeng},
  \citenamefont {Miao}, \citenamefont {Xu}, \citenamefont {Qian}, \citenamefont
  {Weng}, \citenamefont {Richard}, \citenamefont {Fedorov}, \citenamefont
  {Ding}, \citenamefont {Dai},\ and\ \citenamefont {Fang}}]{fese2015}%
  \BibitemOpen
  \bibfield  {author} {\bibinfo {author} {\bibfnamefont {Z.}~\bibnamefont
  {Wang}}, \bibinfo {author} {\bibfnamefont {P.}~\bibnamefont {Zhang}},
  \bibinfo {author} {\bibfnamefont {G.}~\bibnamefont {Xu}}, \bibinfo {author}
  {\bibfnamefont {L.~K.}\ \bibnamefont {Zeng}}, \bibinfo {author}
  {\bibfnamefont {H.}~\bibnamefont {Miao}}, \bibinfo {author} {\bibfnamefont
  {X.}~\bibnamefont {Xu}}, \bibinfo {author} {\bibfnamefont {T.}~\bibnamefont
  {Qian}}, \bibinfo {author} {\bibfnamefont {H.}~\bibnamefont {Weng}}, \bibinfo
  {author} {\bibfnamefont {P.}~\bibnamefont {Richard}}, \bibinfo {author}
  {\bibfnamefont {A.~V.}\ \bibnamefont {Fedorov}}, \bibinfo {author}
  {\bibfnamefont {H.}~\bibnamefont {Ding}}, \bibinfo {author} {\bibfnamefont
  {X.}~\bibnamefont {Dai}}, \ and\ \bibinfo {author} {\bibfnamefont
  {Z.}~\bibnamefont {Fang}},\ }\href {\doibase 10.1103/PhysRevB.92.115119}
  {\bibfield  {journal} {\bibinfo  {journal} {Phys. Rev. B}\ }\textbf {\bibinfo
  {volume} {92}},\ \bibinfo {pages} {115119} (\bibinfo {year}
  {2015})}\BibitemShut {NoStop}%
\bibitem [{\citenamefont {Nie}\ \emph {et~al.}(2018)\citenamefont {Nie},
  \citenamefont {Xing}, \citenamefont {Jin}, \citenamefont {Xie}, \citenamefont
  {Wang},\ and\ \citenamefont {Prinz}}]{tase2018}%
  \BibitemOpen
  \bibfield  {author} {\bibinfo {author} {\bibfnamefont {S.}~\bibnamefont
  {Nie}}, \bibinfo {author} {\bibfnamefont {L.}~\bibnamefont {Xing}}, \bibinfo
  {author} {\bibfnamefont {R.}~\bibnamefont {Jin}}, \bibinfo {author}
  {\bibfnamefont {W.}~\bibnamefont {Xie}}, \bibinfo {author} {\bibfnamefont
  {Z.}~\bibnamefont {Wang}}, \ and\ \bibinfo {author} {\bibfnamefont {F.~B.}\
  \bibnamefont {Prinz}},\ }\href {\doibase 10.1103/PhysRevB.98.125143}
  {\bibfield  {journal} {\bibinfo  {journal} {Phys. Rev. B}\ }\textbf {\bibinfo
  {volume} {98}},\ \bibinfo {pages} {125143} (\bibinfo {year}
  {2018})}\BibitemShut {NoStop}%
\bibitem [{\citenamefont {Qin}\ \emph {et~al.}(2022)\citenamefont {Qin},
  \citenamefont {Fang}, \citenamefont {Zhang},\ and\ \citenamefont
  {Hu}}]{PhysrevX.12.011030}%
  \BibitemOpen
  \bibfield  {author} {\bibinfo {author} {\bibfnamefont {S.}~\bibnamefont
  {Qin}}, \bibinfo {author} {\bibfnamefont {C.}~\bibnamefont {Fang}}, \bibinfo
  {author} {\bibfnamefont {F.-C.}\ \bibnamefont {Zhang}}, \ and\ \bibinfo
  {author} {\bibfnamefont {J.}~\bibnamefont {Hu}},\ }\href {\doibase
  10.1103/PhysRevX.12.011030} {\bibfield  {journal} {\bibinfo  {journal} {Phys.
  Rev. X}\ }\textbf {\bibinfo {volume} {12}},\ \bibinfo {pages} {011030}
  (\bibinfo {year} {2022})}\BibitemShut {NoStop}%
\end{thebibliography}
%

\clearpage
\begin{widetext}

\section*{Supplementary Materials}

\setcounter{section}{0}
\setcounter{subsection}{0}
\setcounter{equation}{0}
\setcounter{figure}{0}
\setcounter{table}{0}

\renewcommand{\thesubsection}{\Alph{subsection}}
\renewcommand{\theequation}{S\arabic{equation}}
\renewcommand{\thetable}{S\arabic{table}}
\renewcommand{\thefigure}{S\arabic{figure}}

\subsection{\label{sup:A}Calculation methods}
We performed first-principles calculations based on the density functional theory (DFT) using projector augmented wave (PAW) method implemented in the Vienna ab initio simulation package (VASP) \cite{kresse1999} to obtain the electronic structures.
The generalized gradient approximation (GGA), as implemented in the Perdew-Burke-Ernzerhof (PBE) functional \cite{GGA-PBE1996} was adopted.
The cutoff parameter for the wave functions was 520 $\eV$.
The BZ was sampled by Monkhorst-Pack method \cite{Monkhorst-BZ1976} with $4\times 10\times 10$ for the 3D periodic boundary conditions.
The $f$ electrons in lanthanide are neglected in our calculations.

\subsection{\label{sup:B}BSs of the LaAsS- and SrZnSb$_2$- family compounds}
As listed in Table \ref{table:ABX} and the corresponding BSs with SOC shown in Fig. \ref{fig:abx}, there exists a large number of (LaAsS and SrZnSb$_2$ family) compounds sharing similar BSs and same topological nature with the PrAsS system.

\begin{table*}[h!]
    \begin{ruledtabular}
        \caption{LaAsS- and SrZnSb$_2$- family compounds with similar BSs and the same topological nature.}
        \begin{tabular}{c|c c c c c c c c c}
            $(z_2z_2z_2;z_4)$ & Compounds \\\hline
            $(000;2)$         & LaAsS & PrAsS & HoAsS  & ErAsS  & TmAsS  & TbAsS  & DyAsS  & HoAsS  & SrZnSb$_2$ \\
                              & NdAsS & SmAsS & HoAsSe & ErAsSe & TmAsSe & TbAsSe & DyAsSe & HoAsSe &            \\
        \end{tabular}
        \label{table:ABX}
    \end{ruledtabular}
\end{table*}

\begin{figure}[h!]
    \centering
    \includegraphics[width=0.73\textwidth]{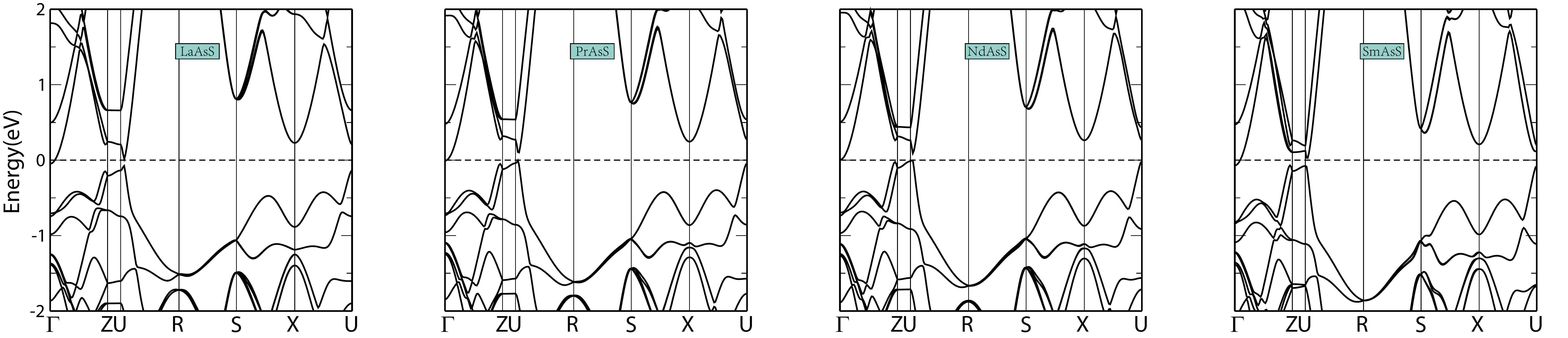}
    \includegraphics[width=0.73\textwidth]{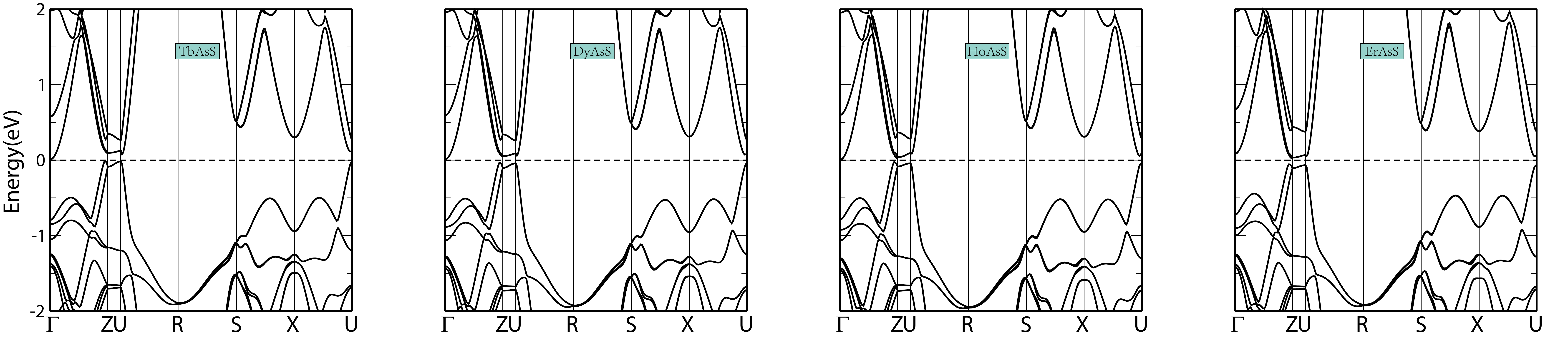}
    \includegraphics[width=0.73\textwidth]{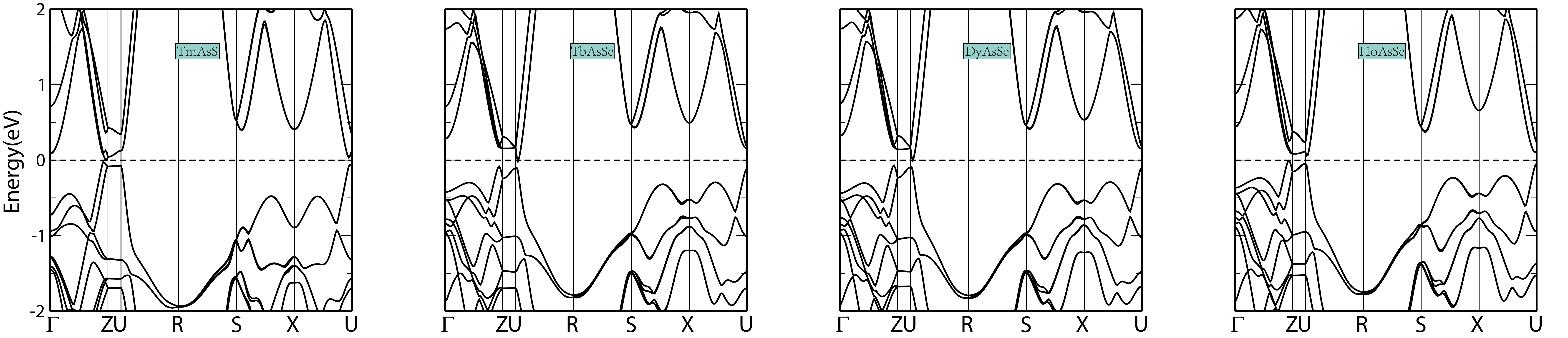}
    \includegraphics[width=0.73\textwidth]{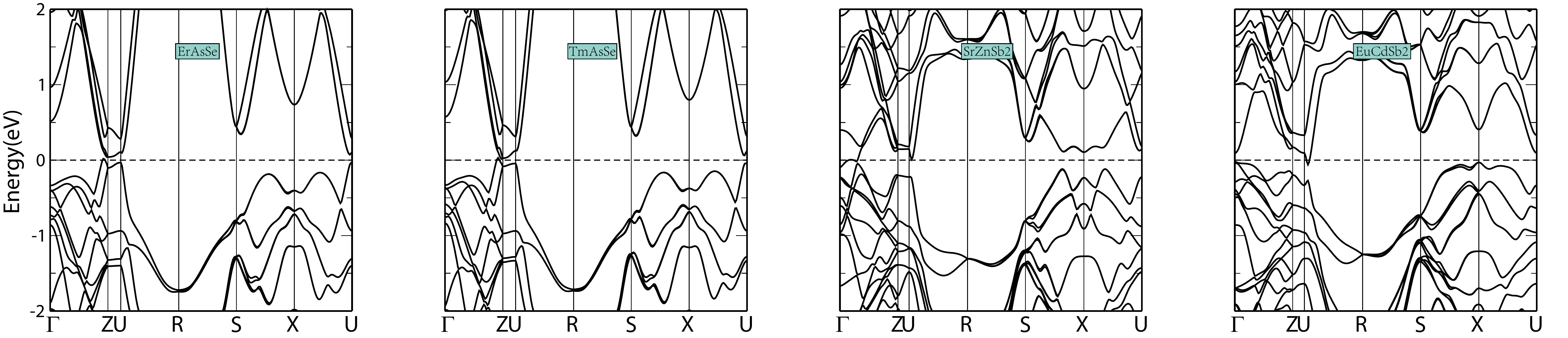}
    \caption{%
        (color online)
        BSs of the LaAsS and SrZnSb$_2$ families with same band topology.
        }
    \label{fig:abx}
\end{figure}

\subsection{\label{sup:C}Phase diagram of the distorted square-net model}
The distorted $X$ square-net monolayer consists of two $X$ atoms locating at $A:\pqty{0, 0}$ and $B: \pqty{1/2-\delta, 1/2}$ in a unit cell (lattice constant $l$).
The parameter $\delta$ describes the distortion, as shown in the main text and Fig. \ref{fig:appc_lat}.
TB model of such a distorted square net is constructed upon the $p_{x,y}$ orbitals of the two $X$ atom.
We take only the nearest-neighboring (NN) hopping (bond 1) and next-nearest-neighboring (NNN) hopping (bond 2 in Fig. \ref{fig:appc_lat}) for simplification in our TB model, which is enough to capture the phase transition between the QSH phase and topologically trivial phase.

\begin{figure}[h!]
    \centering
    \includegraphics[width=0.30\textwidth]{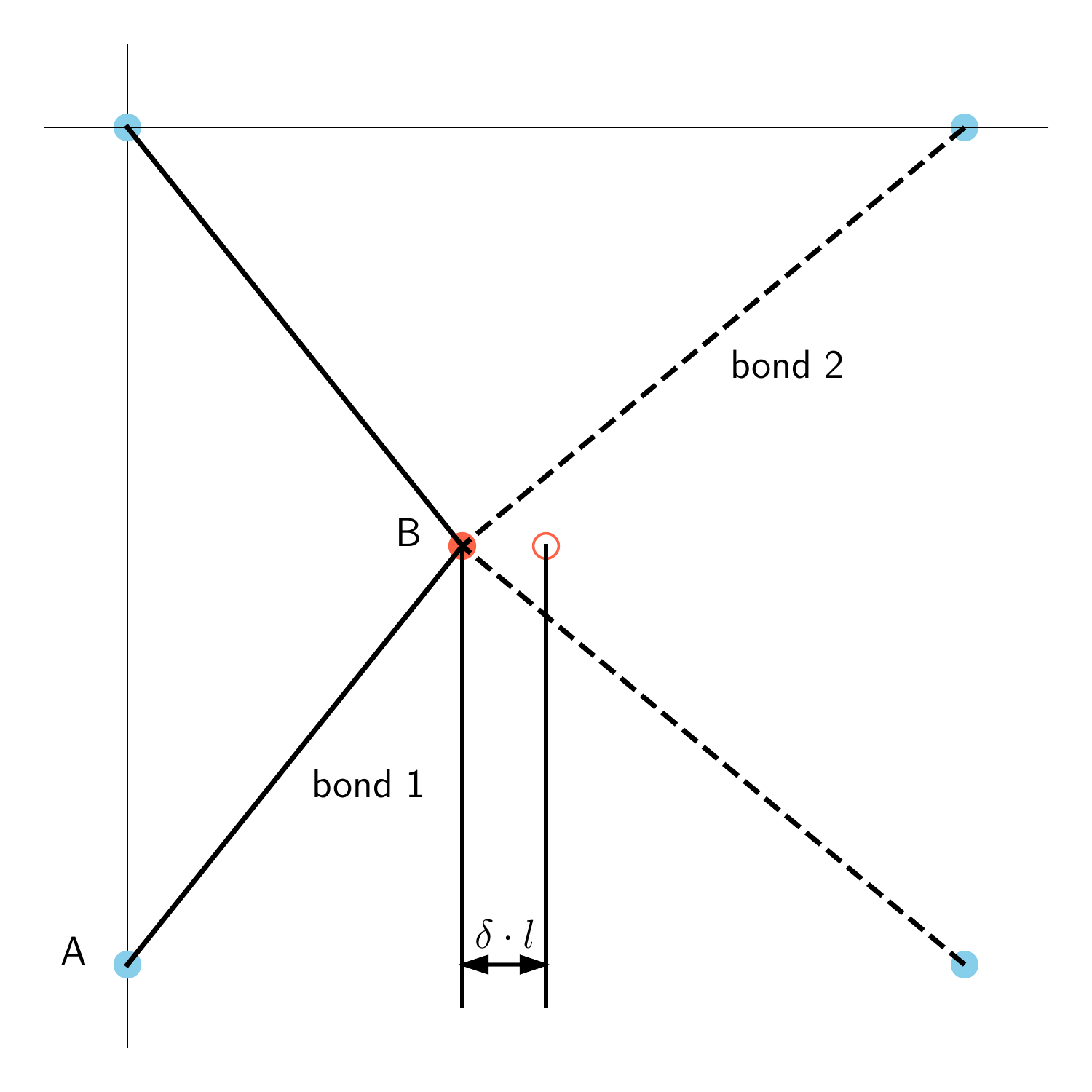}
    \caption{%
        (color online)
        Structure of a distorted $X$ square-net layer, with blue and red cycles representing two sub-lattices, respectively.
        Black solid line denotes the NN hopping (bond 1; $d_1^2 = \bqty{l/2}^2 + \bqty{l/2-\delta\cdot l}^2$), while black dashed line denotes the NNN hopping (bond 2; $d_2^2 = \bqty{l/2}^2 + \bqty{l/2+\delta\cdot l}^2$).
        }
    \label{fig:appc_lat}
\end{figure}

According to the distance ($d$) of the bonds (\ie $d_1$ for bond 1 and $d_2$ for bond 2), we parametrize the Slater-Koster (SK) hopping strength in a uniform way, $V_{i}\pqty{d} \equiv \frac{l^2}{2d^2}V_{i}$, $i\in\Bqty{pp\sigma, pp\pi}$.

First, we introduce the operator
\begin{equation}
    \begin{aligned}
        \psi_{\bk}^{\dg}\equiv & \: \psi_{\bk,\ua}^{\dg}\oplus\psi_{\bk,\da}^{\dg}; \\
        \psi^{\dg}_{\bk, s} \equiv & \: \Bqty{ c^{\dg}_{A, x, s}\pqty{\bk}, c^{\dg}_{A, y, s}\pqty{\bk}, c^{\dg}_{B, x, s}\pqty{\bk}, c^{\dg}_{B, y, s}\pqty{\bk} },
    \end{aligned}
\end{equation}
where $c^{\dg}_{b, \alpha, s}\pqty{\bk}$ is the fermionic creation operator with $b\in\Bqty{A,B}$, $\alpha\in\Bqty{p_x, p_y}$, and $s\in\Bqty{\ua, \da}$, which denotes sub-lattice ($A$ and $B$), orbital ($p_x$ and $p_y$), and spin (up and down), respectively.
The TB Hamiltonian $H_{tb} = H_0 + H_{\text{so}}$ can then be written as (expressed under basis $\psi_{\bk}$, \ie $\Bqty{\ua,\da}\otimes\Bqty{A,B}\otimes\Bqty{p_x,p_y}$)
\begin{equation}
    \begin{aligned}
        H_{tb} = & H_0 + H_{\text{so}} \\
        = & \sum_{\bk} \psi^{\dg}_{\bk} \bqty{ s_0\otimes h_0\pqty{\bk} + h_{\text{so}} } \psi_{\bk}
    \end{aligned}
\end{equation}
Here, the matrix $h_{\text{so}}$ denotes the atomic SOC term of $X$ atoms, which reads,
\begin{equation}
    \begin{aligned}
        h_{\text{so}} = \lambda_{\text{so}}\: s_z \otimes \tau_0 \otimes \sigma_y
    \end{aligned}
\end{equation}
with $\lambda_{\text{so}}$ the strength of SOC, while $\bs$, $\btau$ and $\bsig$ are Pauli matrices in spin, sub-lattice, and orbital space, respectively.
Note that $s_{z}$ is conserved in this model due to the $M_{z}$ symmetry in the basis of $p_{x}$ and $p_{y}$.
While the Hermitian matrix $h_0\pqty{\bk}$ can be expressed as
\begin{equation}
    \begin{aligned}
        h_{0}\pqty{\bk} = &
        \begin{pmatrix}
            & & h_{13}\pqty{\bk} & h_{14}\pqty{\bk} \\
            & & h_{23}\pqty{\bk} & h_{24}\pqty{\bk} \\
            \dg & \dg \\
            \dg & \dg
        \end{pmatrix}
    \end{aligned}
\end{equation}
where the matrix elements $h_{\beta \beta'}$, $\beta\in\Bqty{1,2,3,4}\equiv\Bqty{A,B}\otimes\Bqty{p_x,p_y}$, are given by
\begin{equation}
    \begin{aligned}
        h_{13}\pqty{\bk} = & \pqty{ 2 t^{13}_{1}e^{i \pqty{\frac12-\delta} k_x} + 2 t^{13}_{2} e^{-i \pqty{\frac12+\delta} k_x} } \cos{\frac{k_y}{2}}, \\
        h_{24}\pqty{\bk} = & \pqty{ 2 t^{24}_{1}e^{i \pqty{\frac12-\delta} k_x} + 2 t^{24}_{2} e^{-i \pqty{\frac12+\delta} k_x} } \cos{\frac{k_y}{2}}, \\
        h_{14}\pqty{\bk} = h_{23}\pqty{\bk}
        = & \pqty{ 2i t^{14}_{1}e^{i \pqty{\frac12-\delta} k_x} + 2i t^{14}_{2} e^{-i \pqty{\frac12+\delta} k_x} } \sin{\frac{k_y}{2}}, \\
    \end{aligned}
\end{equation}
Here, $t^{\beta \beta'}_{j}$, $j\in\Bqty{1,2}$ corresponds to two kinds of bonds, which can be obtained via SK parameters extracted from first-principle calculations.
\begin{equation}
    \begin{aligned}
        & t^{13}_{1} = \pqty{\frac{l/2}{d_1}}^{2} V_{pp\pi}\pqty{d_1} + \pqty{\frac{l/2-\delta l}{d_1}}^{2} V_{pp\sigma}\pqty{d_1} \\
        & t^{13}_{2} = \pqty{\frac{l/2}{d_2}}^{2} V_{pp\pi}\pqty{d_2} + \pqty{\frac{l/2+\delta l}{d_2}}^{2} V_{pp\sigma}\pqty{d_2} \\
        & t^{24}_{1} = \pqty{\frac{l/2}{d_1}}^{2} V_{pp\sigma}\pqty{d_1} + \pqty{\frac{l/2-\delta l}{d_1}}^{2} V_{pp\pi}\pqty{d_1} \\
        & t^{24}_{2} = \pqty{\frac{l/2}{d_2}}^{2} V_{pp\sigma}\pqty{d_2} + \pqty{\frac{l/2+\delta l}{d_2}}^{2} V_{pp\pi}\pqty{d_2} \\
        & t^{14}_{1} = \frac{l/2}{d_1}\frac{l/2-\delta l}{d_1} \bqty{V_{pp\sigma}\pqty{d_1} - V_{pp\pi}\pqty{d_1}} \\
        & t^{14}_{2} = \frac{l/2}{d_2}\frac{l/2+\delta l}{d_2} \bqty{V_{pp\sigma}\pqty{d_2} - V_{pp\pi}\pqty{d_2}} \\
    \end{aligned}
\end{equation}

\subsection{\label{sup:D}Minimum tight-binding model for the distorted compounds in SG 62}
According to the projected BSs shown in Fig.2(c,d) in the main text, we find that bands near $E_F$ are mainly contributed by the As-$p_{y,z}$ orbitals.
Furthermore, the band inversion around $U$ which brings the nontrivial topology can be well depicted within these states.
Thus, we can simplify our analysis within a sixteen-band TB model based on 8 spinfull orbitals (As-$p_{y,z}$) of 4 As atoms,
locating at $4c(m)$ Wyckoff positions (WKPs) of SG 62, with reduced coordinates As(A): $\pqty{x, 1/4, z}$, As(B): $\pqty{1-x, 3/4, 1-z}$, As(C): $\pqty{-x+1/2, 3/4, z+1/2}$, and As(D): $\pqty{x+1/2, 1/4, -z+1/2}$ [$x=0.0008$ (slightly buckled) and $z=0.2254$ in the material], the crystal structure viewing parallel to axis $\hbb$ is shown in Fig. \ref{fig:s4_bond}.
Parameters $x$ and $z$ can be viewed as small distortions along $\hba$ and $\hbc$ directions respectively.
Only 5 kinds of bonds between As atoms have been considered in this model, denoted as $E_i$, $i\in \Bqty{1,2,...,5}$, as shown in Fig. \ref{fig:s4_bond}, $E_{1}$, $E_{2}$ bonds are intra-layer hoppings, while $E_{3}$, $E_{4}$, $E_{5}$ bonds are inter-layer hoppings.

\begin{figure}[h!]
    \centering
    \includegraphics[width=0.45\textwidth]{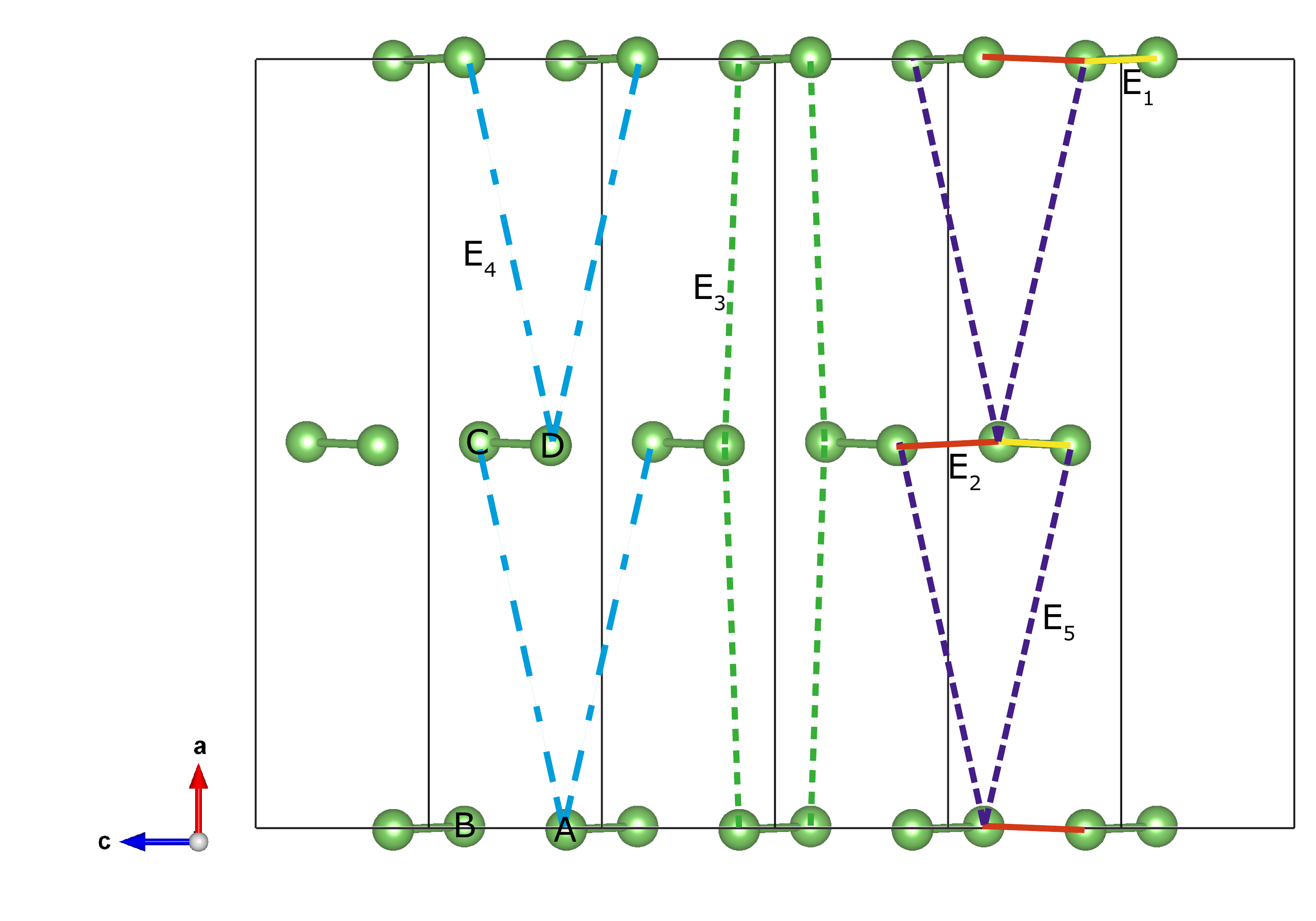}
    \caption{%
        (color online)
        Lattice structure of the TB model and 5 kinds of bonds (
        $E_{1}$: yellow solid lines,
        $E_{2}$: red solid lines,
        $E_{3}$: green dashed lines,
        $E_{4}$: blue dashed lines,
        $E_{5}$: purple dashed lines).
        }
    \label{fig:s4_bond}
\end{figure}

\begin{table}[!h]
    \caption{%
        The SK parameters of $h_{0}'(\bk)$ in Eq. \ref{eq:hamk.62} for the $Pnma$ lattice, with atoms locating at $4c(m)$ WKP.
        }
    \begin{tabular}{c|c|c|c|c|c} \hline\hline
        $i$                 & 1 & 2 & 3 & 4 & 5 \\\hline
        $V_{pp\sigma}^{i}$  & $ 2.09$ & $ 1.53$ & $ 0.00936$ & $ 0.00807$ & $ 0.00726$ \\\hline
        $V_{pp\pi}^{i}$     & $-0.77$ & $-0.54$ & $-0.00187$ & $-0.00161$ & $-0.00145$ \\\hline\hline
    \end{tabular}
    \label{table:bl-net.lda}
\end{table}

\begin{table}[!h]
    \caption{
        The hopping parameters of $h_{\text{so},1}'(\bk)$ in Eq. \ref{eq:hamk.62} for a SG 62 lattice, with atoms locating at $4c(m)$ WKP.
    }
    \begin{tabular}{c|c|c|c|c|c}\hline\hline
        $i$                       & 1 & 2 & 3 & 4 & 5 \\\hline
        $\lambda^{i}_{yy,\ua\da}$ & & & $0.003i$ & $0.04 + 0.02i$ & $0.03 + 0.01i$ \\\hline
        $\lambda^{i}_{yz,\ua\da}$ & $0.08i$ & $0.07i$ & $0.001i$ & $0.05 + 0.03i$ & $0.04 + 0.02i$ \\\hline
        $\lambda^{i}_{zz,\ua\da}$ & & & $0.002i$ & $0.07 + 0.05i$ & $0.06 + 0.04i$ \\\hline\hline
    \end{tabular}
    \label{table:bl-net.soc}
\end{table}

Thus, we construct the basis as below,
\begin{equation}
    \begin{aligned}
        {\psi_{\bk}'}^{\dg} \equiv & \: {\psi_{\bk,\ua}'}^{\dg}\oplus{\psi_{\bk,\da}'}^{\dg}; \quad
        {\psi_{\bk,s}'}^{\dg} \equiv & \: \Bqty{ c^{\dg}_{A,y,s}\pqty{\bk}, c^{\dg}_{A,z,s}\pqty{\bk}, ..., c^{\dg}_{D,y,s}\pqty{\bk}, c^{\dg}_{D,z,s}\pqty{\bk} }
    \end{aligned}
\end{equation}
where $c^{\dg}_{b, \alpha, s}\pqty{\bk}$ is the fermionic creation operator with $b\in\Bqty{A,B,C,D}$, $\alpha\in\Bqty{p_y, p_z}$, and $s\in\Bqty{\ua,\da}$ being sub-lattice, orbital, and spin indices, respectively.
Therefore, the TB Hamiltonian $H_{tb}'$ can be written as (expanded under basis $\psi_{\bk}'$, \ie $\Bqty{\ua,\da}\otimes\Bqty{A,B,C,D}\otimes\Bqty{p_y,p_z}$)
\begin{equation}\label{eq:hamk.62}
    \begin{aligned}
        H_{tb}' = & \; \sum_{\bk} {\psi_{\bk}'}^{\dg} \bqty{ s_{0} \otimes h_{0}'\pqty{\bk} + h_{\text{so},0}' + h_{\text{so},1}'\pqty{\bk} } \psi_{\bk}'
    \end{aligned}
\end{equation}
in which
$h_{0}'(\bk)$ is the SK hopping term,
$h_{\text{so},0}' = \lambda_{\text{so}}'\: s_x \otimes \bbI_{4\times4} \otimes \sigma_y$ is the atomic SOC term, with $\lambda_{\text{so}}'=0.08\eV$ the atomic SOC strength, $\bbI_{4\times4}$ the identity matrix in sub-lattice space, $\bs$ and $\bsig$ the Pauli matrices in spin and orbital space, respectively.
The SK parameters for $h_{0}'(\bk)$ are listed in Table \ref{table:bl-net.lda},
and the hopping parameters $\lambda^{i}_{\alpha\alpha',\ua\da}$ for $h_{\text{so},1}'(\bk)$ are listed in Table \ref{table:bl-net.soc}, defined as below
\begin{equation}
    \begin{aligned}
        \lambda^{1}_{\alpha\alpha', \ua\da} = & \;
        \mel{(000),A,\alpha,\ua}{H}{(\bar{1}\bar{1}\bar{1}),B,\alpha',\da} \\
        \lambda^{2}_{\alpha\alpha', \ua\da} = & \;
        \mel{(000),A,\alpha,\ua}{H}{(\bar{1}00),B,\alpha',\da} \\
        \lambda^{3}_{\alpha\alpha', \ua\da} = & \;
        \mel{(000),A,\alpha,\ua}{H}{(000),D,\alpha',\da} \\
        \lambda^{4}_{\alpha\alpha', \ua\da} = & \;
        \mel{(000),A,\alpha,\ua}{H}{(\bar{1}00),C,\alpha',\da} \\
        \lambda^{5}_{\alpha\alpha', \ua\da} = & \;
        \mel{(000),A,\alpha,\ua}{H}{(000),C,\alpha',\da}
    \end{aligned}
\end{equation}


Hopping terms included in $h_{\text{so},1}'(\bk)$ are symmetrized to and can be derived from those listed in Table \ref{table:bl-net.soc}, with respect to generators (\ie $\tilde{C}_{2z}$, $\tilde{C}_{2y}$ and $\calI$) of SG 62 and time-reversal (TR) operation $\calT$.
In general, we can label an arbitrary As atom by $\pqty{ijk;b}$, with Miller indices $(ijk)$ and $b\in\Bqty{A,B,C,D}$.
Any symmetry operator $R$ in SG 62 maps site $\pqty{ijk;b}$ to $\pqty{i'j'k';b'}$, \ie $R: \pqty{ijk;b}\mto\pqty{i'j'k';b'}$.
The concrete representations of these operators in orbital ($\alpha$) and spin ($s$) space,
\ie $R: c^{\dg}_{\alpha(s)}\mto\sum_{\alpha'(s')} c^{\dg}_{\alpha'(s')}\mU(R)_{\alpha'\alpha(s's)}$, are listed in Table \ref{table:rep62}.

\begin{table*}[b!]
    \caption{
        Representations of SG 62 generators and TR operation
    }
    \begin{tabular}{c|c|c|c|c}\hline\hline
        & $\mathcal T$ & $\tilde{C}_{2z}$ & $\tilde{C}_{2y}$ & $\mathcal I$ \\\hline
        $\pqty{ijk;b}$ & $\pqty{ijk;b}$ &
        $\mqty{\pqty{ijk;A}\mto\pqty{\bar{i}\bar{j}k;C}+\pqty{\bar{1}\bar{1}0} \\
               \pqty{ijk;B}\mto\pqty{\bar{i}\bar{j}k;D}+\pqty{\bar{1}\bar{1}\bar{1}} \\
               \pqty{ijk;C}\mto\pqty{\bar{i}\bar{j}k;A}+\pqty{\bar{1}\bar{1}\bar{1}} \\
               \pqty{ijk;D}\mto\pqty{\bar{i}\bar{j}k;B}+\pqty{\bar{1}\bar{1}0} }$ &
        $\mqty{\pqty{ijk;A}\mto\pqty{\bar{i}j\bar{k};B}+\pqty{\bar{1}0\bar{1}} \\
               \pqty{ijk;B}\mto\pqty{\bar{i}j\bar{k};A}+\pqty{\bar{1}1\bar{1}} \\
               \pqty{ijk;C}\mto\pqty{\bar{i}j\bar{k};D}+\pqty{\bar{1}1\bar{1}} \\
               \pqty{ijk;D}\mto\pqty{\bar{i}j\bar{k};C}+\pqty{\bar{1}0\bar{1}} }$ &
        $\mqty{\pqty{ijk;A}\mto\pqty{\bar{i}\bar{j}\bar{k};B}+\pqty{\bar{1}\bar{1}\bar{1}} \\
               \pqty{ijk;B}\mto\pqty{\bar{i}\bar{j}\bar{k};A}+\pqty{\bar{1}\bar{1}\bar{1}} \\
               \pqty{ijk;C}\mto\pqty{\bar{i}\bar{j}\bar{k};D}+\pqty{\bar{1}\bar{1}\bar{1}} \\
               \pqty{ijk;D}\mto\pqty{\bar{i}\bar{j}\bar{k};C}+\pqty{\bar{1}\bar{1}\bar{1}} }$ \\\hline
        $\alpha$ & $\mqty(1&0\\0&1)$ & $\mqty(-1&0\\0&1)$ & $\mqty(1&0\\0&-1)$ & $\mqty(-1&0\\0&-1)$ \\\hline
        $s$      & $\mqty(0&-1\\1&0)$ & $\mqty(-i&0\\0&i)$ & $\mqty(0&-1\\1&0)$ & $\mqty(1&0\\0&1)$ \\\hline\hline
    \end{tabular}
    \label{table:rep62}
\end{table*}

\subsection{\label{sup:E}The Wilson-loop spectrum for $M_y$ MCNs and the hourglass invariant $g_z$}
To characterize the topological properties in the PrAsS system, $M_y$ MCNs $(m_0,m_\pi)$ and the hourglass invariant $g_z$ are calculated by the Wilson-loop method.
As shown in Fig. \ref{fig:s5-my}(a) and \ref{fig:s5-my}(b), $M_y$ MCNs are calculated to be (0,0) in $k_{y}=(0,\pi/b)$ planes.
Similarly, $g_z$ invariant and SCNs in $k_{x}=(0,\pi/a)$ planes are calculated to be $1$ [Fig. \ref{fig:s5-my}(c) and \ref{fig:s5-my}(d)] and $(2,2)$ [Fig. 4(a) and 4(b) in the main text].

\begin{figure}[h!]
    \centering
    \includegraphics[width=0.49\textwidth]{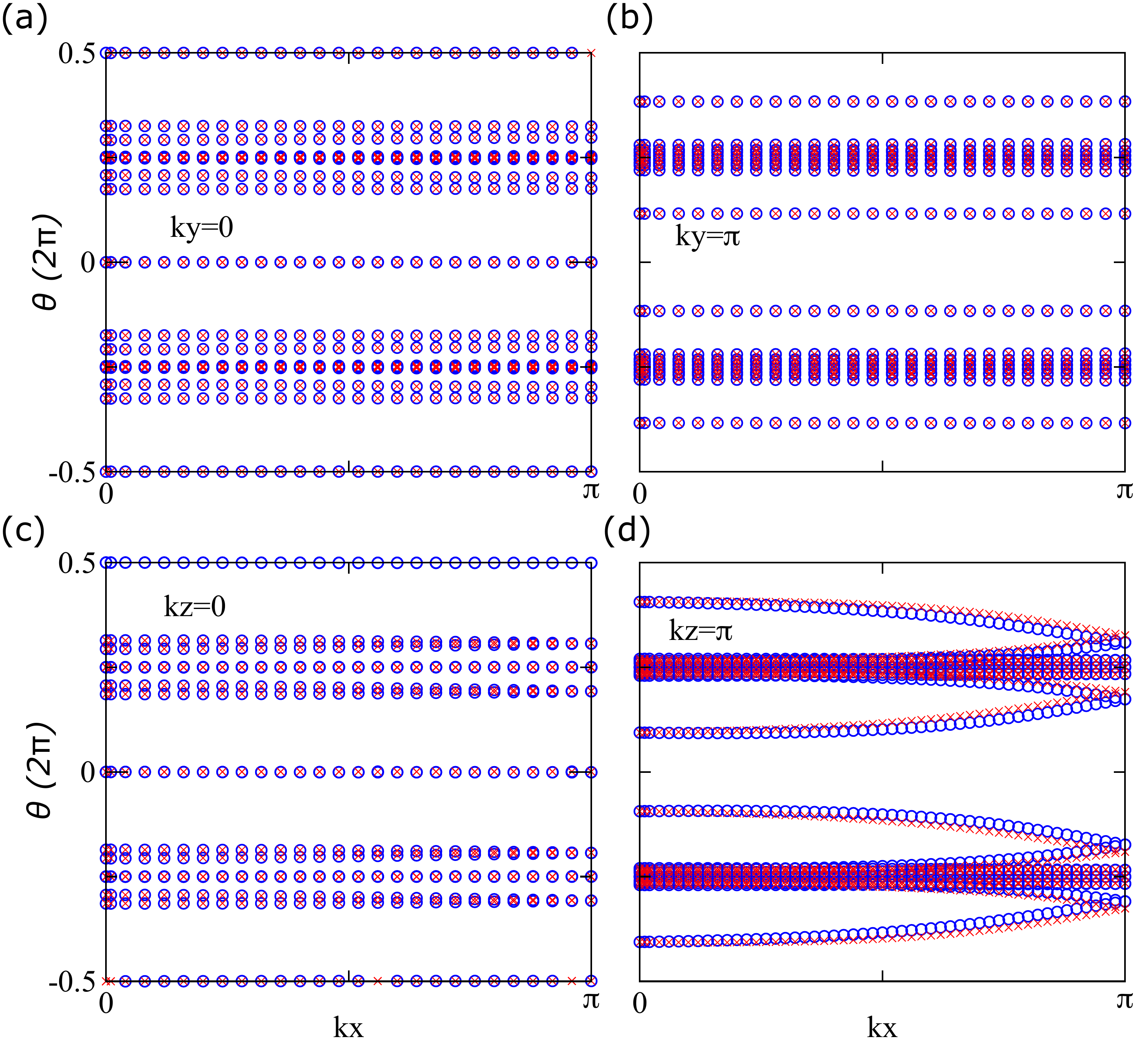}
    \caption{%
        (color online)
        $k_z$- and $k_y$- directed Wilson-loop spectra in bulk PrAsS.
        MCNs at (a) $k_y=0$ plane and (b) $k_y=\pi/b$ plane.
        $g_z$, \ie the hourglass-shaped evolution of $k_z$-directed Wilson-loop eigenvalues on the (c) $\bar{\Gamma}-{\rm\bar{X}}$ line and (d) ${\rm\bar{U}-\bar{Z}}$ line.
        }
    \label{fig:s5-my}
\end{figure}

\clearpage
\subsection{\label{sup:F}DFT bands for $X$ square-net monolayers}
Using SK parameters listed in Table \ref{table:Xnet}, BSs from the TB model reproduce the $p_{x,y}$-resolved BSs of $X$ square-net monolayers with and without SOC well near $E_F$.
The TB model is constructed on an $X$ square net with $p_{x,y}$ orbitals shown as the inset of Fig. \ref{fig:s6_nso}(a).
Only the nearest-neighboring (NN) $X$--$X$ hopping terms and the next-nearest-neighboring (NNN) $X$--$X$ hopping terms are considered in this model.
\begin{table*}[h!]
    \begin{ruledtabular}
        \caption{%
            The fitted SK parameters from the first-principles calculations for an $X$ square-net layer (lattice constant $l$).
            The SK parameters for the NN $X$--$X$ bonds are $V_{pp\sigma,pp\pi}$,
            and those of the NNN $X$--$X$ bonds are $V_{pp\sigma,pp\pi}'$.
            }
        \begin{tabular}{l c c c c c l}
            X    & $l$ (\AA) & $V_{pp\sigma},V_{pp\pi}$ (eV) & $V_{pp\sigma}',V_{pp\pi}'$ (eV) & $V_{pp\sigma}/V_{pp\pi}$ & $\lambda_{so}$ (eV) & Example \\\hline
            P    &  3.8      & -2.0,  0.700  & -0.40,  0.1400  &  -0.35  &  0.020  &  GdPS        \\
            As   &  3.9      & -2.0,  0.660  & -0.30,  0.0990  &  -0.33  &  0.105  &  PrAsS       \\
            Sb   &  4.4      & -1.8,  0.576  & -0.25,  0.0800  &  -0.32  &  0.230  &  EuCdSb$_2$  \\
            Bi   &  4.5      & -1.8,  0.504  & -0.01,  0.0028  &  -0.28  &  0.770  &  CaMnBi$_2$  \\
        \end{tabular}
        \label{table:Xnet}
    \end{ruledtabular}
\end{table*}

\begin{figure}[h!]
    \centering
    \includegraphics[width=.98\textwidth]{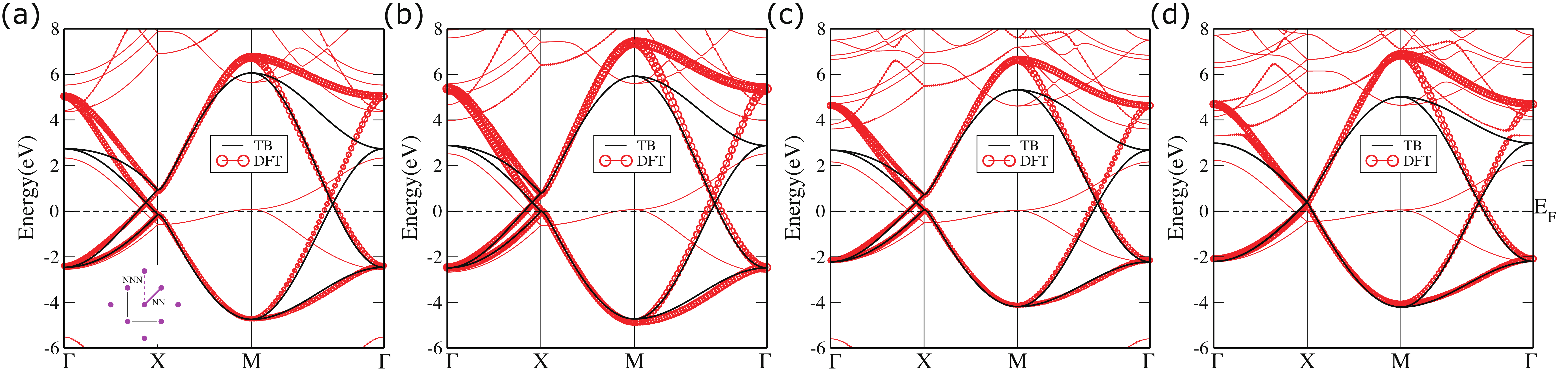}
    \caption{%
        (color online)
        The orbital-resolved BSs of $X$ square-net monolayers without SOC from DFT (labeled by ``DFT" in the legend), the size of red cycles represents the weight of $X$-$p_{x,y}$ orbitals. The BSs of fitted TB model is shown by black lines.
        The inset of (a) is the unit cell of an $X$ square net with solid (dashed) line being the NN (NNN) bond.
        }
    \label{fig:s6_nso}
\end{figure}
\begin{figure}[h!]
    \centering
    \includegraphics[width=.98\textwidth]{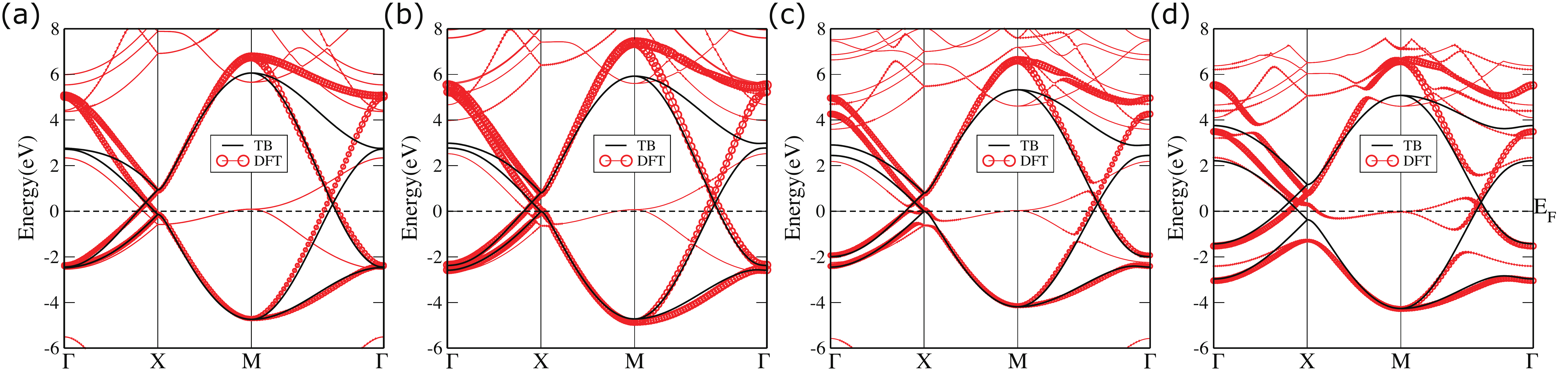}
    \caption{%
        (color online)
        The orbital-resolved BSs of $X$ square-net monolayers with SOC from DFT (labeled by ``DFT" in the legend), the size of red cycles represents the weight of $X$-$p_{x,y}$ orbitals. The BSs of fitted TB model is shown by black lines.
        Only onsite $h_{so} = \lambda_{so} s_{z}\otimes \tau_{0}\otimes \sigma_{y}$ is considered in the SOC TB model
        }
    \label{fig:s6_soc}
\end{figure}
\clearpage
\end{widetext}
\end{document}